\documentclass{aa}  

\usepackage{graphicx}
\usepackage{txfonts}
\begin{document}

   \title{Bulge formation inside quiescent lopsided stellar disks: connecting accretion, star formation and morphological transformation in a $\rm z\sim3$ galaxy group}


   \author{ Boris S. Kalita\inst{1}, Emanuele Daddi\inst{1} \and Frederic Bournaud\inst{1} \and R.~Michael~Rich\inst{2} \and Francesco Valentino\inst{3,4} \and Carlos G\'omez-Guijarro\inst{1} \and Sandrine Codis\inst{1} \and Ivan Delvecchio\inst{5} \and David Elbaz\inst{1} \and Veronica Strazzullo\inst{6,7,8} \and Victor de Sousa Magalhaes\inst{8} \and J\'er\^ome Pety\inst{9} \and Qinghua Tan\inst{10,1} 
     } 

   \institute{
CEA, Irfu, DAp, AIM, Universit\`e Paris-Saclay, Universit\`e de Paris, CNRS, F-91191 Gif-sur-Yvette, France
\and Department of Physics \& Astronomy, University of California Los Angeles, 430 Portola Plaza, Los Angeles, CA 90095, USA
\and Cosmic Dawn Center (DAWN) 
\and Niels Bohr Institute, University of Copenhagen, Jagtvej 128, DK-2200, Copenhagen N, Denmark
\and INAF — Osservatorio Astronomico di Brera, via Brera 28, I-20121, Milano, Italy 
\and University of Trieste, Piazzale Europa, 1, 34127 Trieste TS, Italy
\and INAF — Osservatorio Astronomico di Brera, via Brera 28, I-20121, Milano, Italy
\and INAF — Osservatorio Astronomico di Trieste, via Tiepolo 11, I-34131, Trieste, Italy
\and Institut de Radioastronomie Millim\'etrique, 300 Rue de la Piscine, F-38406 Saint Martin d’H\`eres, France
\and Purple Mountain Observatory \& Key Laboratory for Radio Astronomy, Chinese Academy of Sciences, 10 Yuanhua Road, Nanjing 210023, People's Republic of China 
 } 
\offprints{Boris S. Kalita \\
\email{boris.kalita@cea.fr}}
   \date{\centering Accepted for publication in A\&A}

 
  \abstract{
We present well-resolved near-IR and sub-mm analysis of the three highly star-forming massive ($>10^{11}\,\rm M_{\odot}$) galaxies within the core of the RO-1001 galaxy group at $\rm z=2.91$. Each of them displays kpc-scale compact star-bursting cores with properties consistent with forming galaxy bulges, embedded at the center of extended, massive stellar disks. Surprisingly, the stellar disks are unambiguously both quiescent, and severely lopsided. Therefore,  `outside-in' quenching is ongoing in the three group galaxies.
We propose an overall scenario in which the strong mass lopsidedness in the disks (ranging from factors of 1.6 to $>$3), likely generated under the effects of accreted gas and clumps, is responsible for their  star-formation suppression, while funnelling gas into the nuclei and thus creating the central starbursts. The lopsided side of the disks marks the location of accretion streams impact, with additional matter components (dust and stars) detected in their close proximity directly tracing the inflow direction. The interaction with the accreted clumps, which can be regarded as minor-mergers, leads the major axes of the three galaxies to be closely aligned with the outer Lyman-$\alpha$-emitting feeding filaments. These results provide the first observational evidence of the impact of  cold accretion streams on the formation and evolution of the galaxies they feed. In the current phase, this is taking the form of the rapid buildup of bulges under the effects of accretion, while still preserving  massive quiescent and lopsided stellar disks at least until encountering a violent major-merger.

} 

\titlerunning{Bulge formation inside quiescent lopsided stellar disks}
\authorrunning{Kalita et al.}
\maketitle
%

\section{Introduction}

The formation of massive galaxies ($> 10^{11}\,\rm M_{\odot}$) that happened within the first $\sim 3\,\rm Gyr$ after the Big Bang sustained by cold streams \citep{birmboim03,keres05}, 
is still not fully understood \citep{somerville15, vogelsberger20}.
Cold gas accretion (which includes smooth gas as well as clumpy material) is expected to be ubiquitous in high redshift dense environments \citep{dekel09a}. Encouraging   observational evidence for cold streams was recently obtained for a sample of galaxy groups and clusters at $2.0 < \rm z < 3.3$ and for statistical samples of massive galaxies \citep{daddi22b, daddi22a}. Yet there remains a paucity of direct observations of how this critical process  affects individual galaxies. In general, the need of gas accretion fueling the star formation in galaxies is abundantly clear, exemplified by the tight correlation between their star formation rates (SFR) and the stellar mass at least up to $\rm z \sim 4-6$ \citep[the star-forming main-sequence;][]{daddi07, elbaz07, whitaker12, speagle14, schreiber15}. Morphologically, this phase is expected to feature secularly growing disks \citep{daddi10, tacconi10, dekel13b, feldmann15}. 

Meanwhile, to accomodate the existence of compact ($\sim 1\,\rm kpc$), spheroidal, quiescent galaxies (QGs) observed at $\rm z \sim 2-4$ \citep[e.g.,][]{valentino20, lustig21, deugenio21} in contrast to the relatively more extended star-forming disk galaxies \citep{vanderwel14}, an intermediate population of compact star-forming galaxies (cSFGs) have been proposed and widely observed \citep{barro13, barro14, vandokkum15, puglisi19, puglisi21} . The feature at the heart of this hypothesis, quite literally, is the compact star-forming core that is expected to form the spheroidal bulge characteristic of the compact QGs. When gas is driven into the central compact region, the increase in the gas surface density results in a rise in star-formation that rapidly builds up the stellar mass of the bulge. However, the mode of formation of these cSFGs has been a matter of debate. On one hand, simulations predict that smooth-gas-accretion driven violent disk instabilities result in the gas being driven to the core of secularly evolving disk-like galaxies, once the galaxy mass reaches $\sim 10^{9.5}\,\rm M_{\odot}$ and thereby creating most of the stellar mass in a compact spheroidal structure \citep[\emph{wet compaction};][]{dekel13, zolotov15, tacchella16}. Whereas observational studies of large statistical samples advocate the rapid effects of major-mergers between galaxies leading to the formation of such objects \citep{cimatti08, ricciardelli10, fu13, ivison13, toft14, toft17, elbaz18, gomez-guijarro18}. Both scenarios have its issues however. Wet compaction predicts the presence of high-gas fractions in cSFGs that is simply not observed \citep{puglisi21, gomez-guijarro21, gomez-guijarro22}. Whereas, a dominance of the merger-driven starbursts (pushing them a factor of $3-4$ above the main sequence) would suggest that the galaxy simply transitions through the star-forming main-sequence on its way to becoming quiescent, making the tight correlation between the SFR and stellar masses difficult to reproduce. Although, \citet{elbaz18} does provide an tentative solution through a sub-population of star-bursts `hidden' within the main-sequence. 

The morphology of cSFGs could hold clues to disentangling the processes. 
Besides their characteristic sub-mm bright highly star-forming cores, the surrounding stellar regions can also provide valuable information, which is only possible through near-IR follow-ups of ALMA (sub-mm) detected cSFGs. In the wet compaction scenario, once the galaxy enters its peak compaction phase at $\sim 10^{9.5}\,\rm M_{\odot}$, the stellar disk (or a `proto-disk') stays constant or shrinks while the core rapidly builds up mass \citep{tacchella16}. Whereas in case of major-mergers between massive galaxies, clear signs of disturbed morphology, especially in the form of clumpy stellar structures (at least during the initial phases) is anticipated \citep{lotz06, bournaud11, rujopakarn19, calabro19} and extended tidal-tails \citep[e.g.,][]{bridge10,wen16, guo16}. Attempts at characterizing the stellar regions are already underway \citep[e.g.,][]{puglisi21, gomez-guijarro21}, with the narrative swaying in favour of mergers \citep{elbaz18, rujopakarn19}. Nevertheless, it is imperative to determine whether this is a universal trend, especially in regions where galaxies are expected to grow under the influence of high levels of accretion, which might be more conducive for the wet compaction scenario.  

Therefore, to attempt a resolution, one will have to specifically investigate high-redshift massive dark-matter halos which are expected to allow the penetration and survival of cold-gas streams into a hot halo \citep{keres05, dekel09a}. This is primarily driven by the expected evolution with halo mass ($\propto \rm M_{DM}$) and redshift  ($\propto \rm (1+z)^{\alpha}$ with $\alpha \approx 2.25-2.50$). Studying cSFG counterparts within such accretion-rich environments is our best bet at disentangling the effects of accretion and mergers. However, observationally establishing the presence of infalling cold-gas streams in halos has proved to be extremely challenging, with only a handful of convincing cases yet \citep[using Ly-$\alpha$ emission as a tracer;][]{martin15,martin19, umehata19, daddi21, daddi22b, daddi22a}. Following up on one of these cases, we investigate the galaxy-group RO-1001 at $\rm z = 2.91$ presented in \citet{daddi21}. The structure hosts three massive ($> 10^{11}\,\rm M_{\odot}$) star-forming galaxies in its inner core \citep[along with an additional quiescent galaxy;][]{kalita21b}, at the center of its $\sim 4 \times 10^{13}\,\rm M_{\odot}$ dark-matter halo potential well. They are all $<70$~kpc ($9^{\prime\prime}$) from the peak of the giant 300~kpc-wide Lyman-$\alpha$ halo centered on the group, and spectroscopically coincident with the intensity averaged Lyman-$\alpha$ emission within $\pm 500\,\rm km\,s^{-1}$ (Fig.\ref{fig:RO-1001}). Finally, the Lyman-$\alpha$ contours provide tentative identification  of accretion-streams that are expected to feed the central halo and hence the three star-forming galaxies within.  

It needs to be noted that whenever we refer to accretion in this work, we are including both smooth gas as well as clumpy material. The latter is expected to make up $\sim 1/3$ of the accreted matter \citep{dekel09a} and its interaction with the galaxies could be observationally characterised as minor-mergers of varying mass-ratios. In this paper, we discuss the observational data in Sec.~\ref{sec:obs_data}, followed by the analysis and results (Sec.~\ref{sec:analysis}). We then discuss galaxy morphology, star formation and their link to accretion in Sec.~\ref{sec:disc}. Finally, we provide the conclusions in Sec.\ref{sec:summary}. Throughout, we adopt a concordance $\Lambda$CDM cosmology, characterized by  $\Omega_{m}=0.3$, $\Omega_{\Lambda}=0.7$, and $\rm H_{0}=70$ km s$^{-1}\rm Mpc^{-1}$. We use a Chabrier initial mass function. Magnitudes and colors are on the AB scale. All uncertainties on measured parameters that have been quoted indicate the $90\%$ confidence interval determined from the limits of $\Delta\chi^{2} = 2.71$. All images are oriented such that north is up and east is left.

\section{Observational Data}


\label{sec:obs_data}

To study the three  star-forming galaxies (A, B and C) in RO-1001 (Fig.~\ref{fig:RO-1001}), we make use of \textit{HST/WFC3} imaging in 3 bands (F160W, F125W and F606W) over a total of 11 orbits during Cycle 27 (Proposal ID: 15190, PI: E. Daddi). The data reduction was executed using the pipeline \textit{grizli} (\url{https://github.com/gbrammer/grizli}). The $5 \sigma$ point-source sensitivities reached are 26.25 (F160W), 26.47 (F125W) and 26.39 (F606W) with a pixel scale of $0.06^{\prime\prime}$ and a half-power beam-width of $0.24^{\prime\prime}$ for F160W. Public COSMOS F814W imaging \citep{scoville07b} 
was also incorporated into the analysis. Furthermore, we used the Ks-band image from data release 4 of Ultra-VISTA \citep{mccracken12}. Finally, IRAC $3.6\,\mu$m and $4.5\,\mu$m images were taken from the Spitzer Matching Survey of the Ultra-VISTA Deep Stripes \citep[SMUVS;][]{ashby18}, while those at $5.7\,\mu$m and $7.9\,\mu$m are obtained from the public COSMOS database \citep{laigle16}. 

New Atacama Large Millimetre Array (ALMA) band 7 observations, taken in Cycle 7 (Project ID: 2019.1.00399.S, PI: R.M. Rich) provide the sub-millimeter information of our sources. The data reduction was carried out using the Common Astronomy Software Application (CASA) and a maximum sensitivity of $\rm 28\,\mu Jy\,beam^{-1}$ was reached, with a synthesised beam size of $0.49^{\prime\prime} \times 0.46^{\prime\prime}$. Finally, we utilize NOrthern Extended Millimter Array (NOEMA) spectroscopy data to map the $\rm CO[3-2]$ transition in each of the three galaxies. The synthesized beam size at $88.3\,\rm GHz$ is $4.0^{\prime\prime} \times 1.8^{\prime\prime}$. The spectra creation and analysis is carried out using a GILDAS-based pipeline (\url{http://www.iram.fr/IRAMFR/GILDAS}), and has been already discussed in a previous work \citep{daddi21}.

\section{Analysis}
\label{sec:analysis}


\begin{figure*}
\centering{
\includegraphics[width=\textwidth]{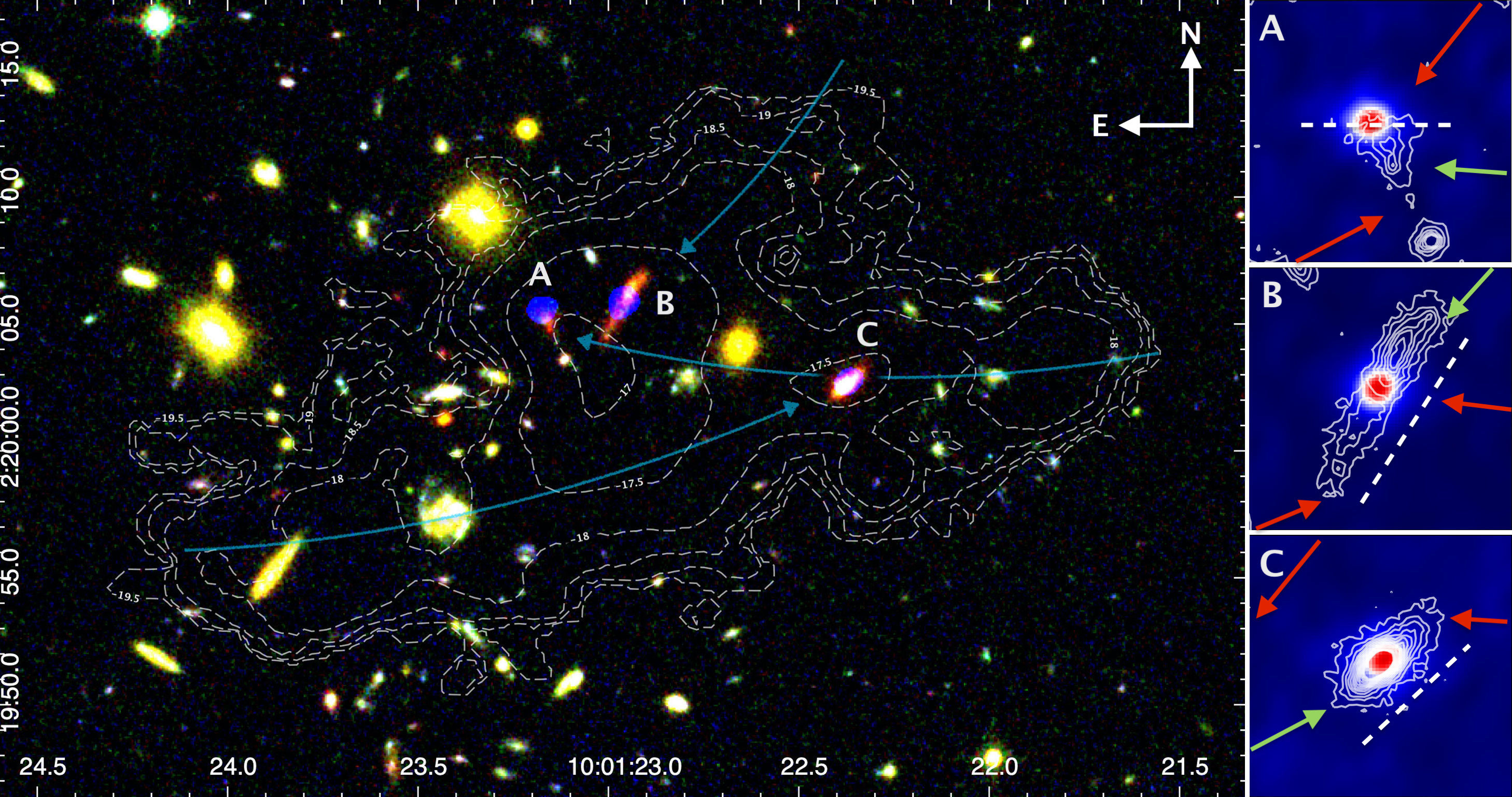}}
\caption{Location of the massive star-forming galaxies in the network of Lyman-$\alpha$ filaments. (Left) Color image composite of RO-1001, using observations in F160W, F125W and F606W. The ALMA data has been co-added to that in F606W since both trace star-formation, differing based on dust obscuration. The scaling of the ALMA data was done based on its flux to SFR ratio in comparison to the same for that in F606W. The white dashed contours trace the Lyman-$\alpha$ halo whereas the cyan arrows show the tentative directions of the three gas-accretion filaments converging onto the center of the group potential well (the correspondence to the galaxies is discussed in detail in Sec.~\ref{sec:lyman_link}). (Right) Zoom on the three star-forming galaxies (A, B and C) as seen in ALMA sub-millimeter dust-emission while the white contours trace the F160W flux tracing stellar light. The contours begin at $4\sigma$ with increments of $4\sigma$. The arrows indicate the direction of the streams, which are shown in green if they are aligned with the major axis of the galactic disks (in dashed white lines), as discussed in Sec.~\ref{sec:lyman_link}. Those which are not, are in red. In case of Galaxy-A, we rather show the major axis of the ALMA contour with the white dashed line.} \label{fig:RO-1001}
\end{figure*}

We use the combination of the data listed above (Sec.~\ref{sec:obs_data}) to derive the properties of the three star-forming galaxies in RO-1001. Each of them show extremely compact sub-millimeter bright rotating cores primarily centered on the much more extended ($4-9\, \times$) near-IR detected stellar components (Fig.\ref{fig:RO-1001}, right panel), with both being co-planar in at least 2/3 cases (Galaxies B and C). The stellar regions also show marked lopsidedness in their emission. We investigate in detail each of these aspects of the galaxies in this section. Given that Galaxy-B is the most extended and hence best resolved, we always begin with the description of its analysis for simplicity. We then proceed to Galaxies A and C, where the same prescription as that for Galaxy-B is used. 

\begin{table*}
{\caption{\centering Properties of the star-forming massive galaxies in RO-1001}
\label{tab:1}
\centering
\begin{tabular}{cllll}
\hline\hline
ID                        &                       & A             & B              & C   \\
\hline
RA                        &                       & 10:01:23.174  & 10:01:22.964   & 10:01:22.369 \\
DEC                       &                       & 02:20:05.57   & 02:20:05.87    & 02:20:02.63 \\
$z_{\rm spec}$            &                       & 2.9214        & 2.9156         & 2.9064 \\
$\log \rm M_{\star}$  (1)       & $\rm M_\odot$             & 11.50$^{+0.15}_{-0.19}$         & 11.20$^{+0.07}_{-0.14}$          & 11.26$^{+0.14}_{-0.10}$  \\
$S_\nu$(870~$\mu$m)       & mJy                   & $4.44\pm0.05$   & $8.69\pm0.03$    & $4.04\pm0.11$\\
SFR$_{ core}$ (2)                   & $\rm M_\odot$~yr$^{-1}$   & $345 \pm 55$           & $674 \pm 106$            & $313 \pm 50$    \\
SFR$_{ disk\_ALMA}$ (2)                   & $\rm M_\odot$~yr$^{-1}$   & $<60^{**}$          & $<19^{**}$            & $<43^{**}$    \\
SFR$_{ disk\_SED}$ (1)                   & $\rm M_\odot$~yr$^{-1}$   & $66^{+446}_{-49}$   (3)        & $42^{+51}_{-37}$            & $94_{-94}^{+66}$    \\
t$_{50}$  (1)               & $\rm Gyr$   & $1.7^{+0.3}_{-0.7}$           & $0.5^{+0.7}_{-0.2}$            & $0.2_{-0.1}^{+1.3}$    \\
r$_{e\_{disk}}$                &  kpc  & $3\pm 1$           & $9 \pm 2$            & $4 \pm 1$    \\
r$_{disk}$/r$_{core}$                &    & $4\pm 1$           & $9 \pm 2$            & $5 \pm 1$    \\
log M$_{mol}$         & M$_{\odot}$ & 9.8 & 10.7  & 10.6\\
FWZV$_{\rm CO[3-2]}$ (4)  & km~s$^{-1}$           & 381           & 1114           & 1098 \\
log M$_{ dyn,tot}$ (5)  & M$_{\odot}$           &     10.4       & 11.1           & 11.2 \\
A$_{v}$  &            &     $1.8_{1.3}^{2.1}$       & $1.0_{0.4}^{1.2}$           & $1.6^{1.8}_{0.4}$ \\

\hline
\hline
\end{tabular}\\
}\vspace{0.2in}
{Notes: (1) Estimated from the composite-$\tau$ model fitting to optical and near-IR photometry and therefore primarily associated with the disks (2) Derived from the $870\mu$m flux of individual galaxies assuming the same SED shape as for their coaddition  \citep{daddi21}. (3) This high upper-limit from the composite-$\tau$ model is in agreement with a constant star formation model being within $90\%$ confidence interval (4) Full Width at Zero Velocity as previously reported \citep{daddi21}. (5) Dynamical mass primarily associated with the core, rather than the whole galaxy as described in the main text.} $^{**}$ $3\sigma$ upper-limits.
\end{table*}

\subsection{{Near-IR surface brightness fitting.}}
\label{sec:morph_stel_fit}

\begin{figure*}[ht]
\centering{
\includegraphics[width=\textwidth]{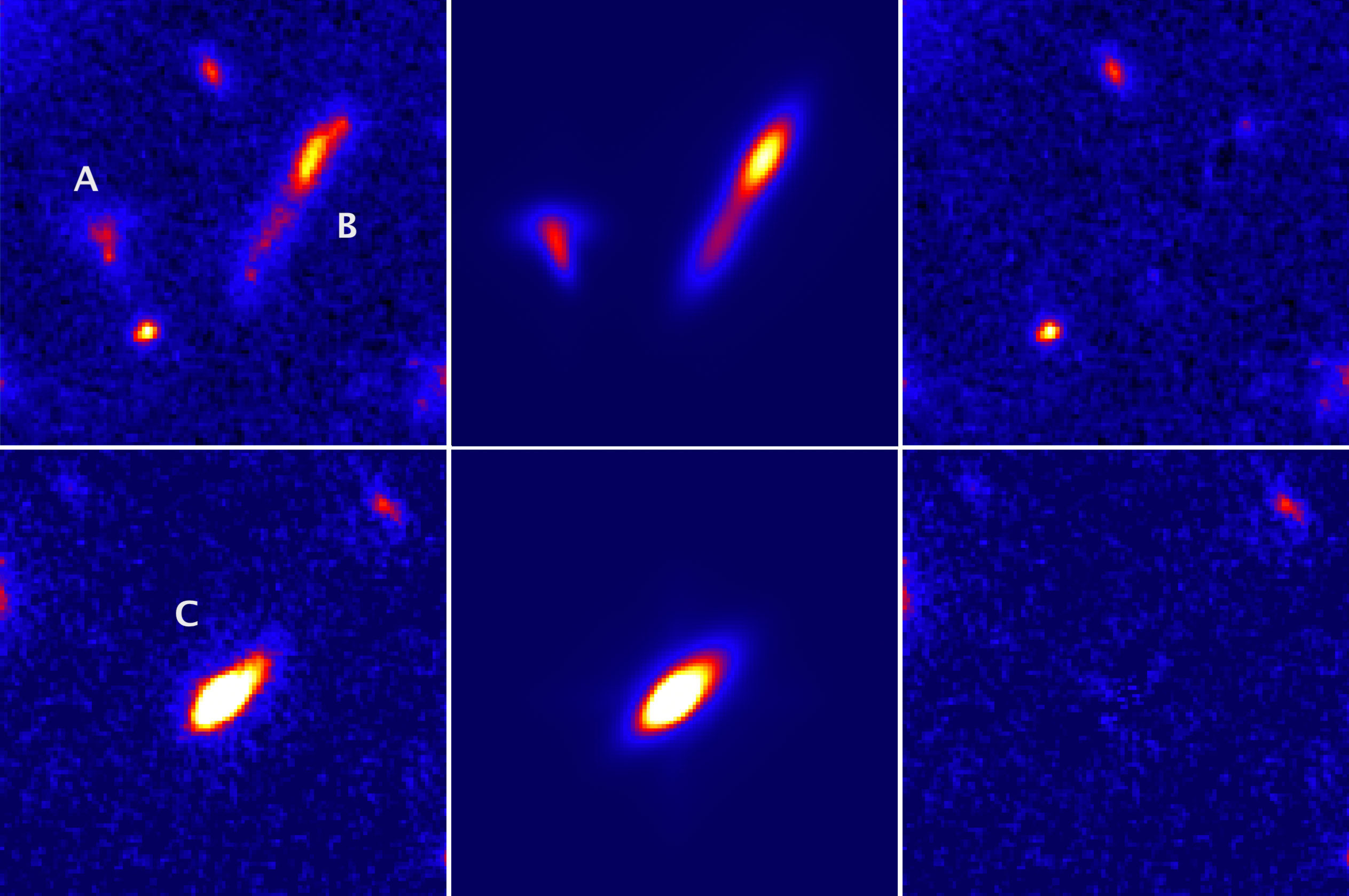}}
\caption{(Left panel) The HST F160W cutouts of Galaxies-A and B (top) and Galaxy-C (bottom) used for the morphological model fitting. (Middle and right panels) The GALFIT returned models and residuals of the same cutouts. In the latter, only the RO-1001 galaxy models have been subtracted. The rest of the sources, although used during the fit, have been left in the image.}\label{fig:galfit_img}
\end{figure*}

We investigate the distribution of rest-frame optical emission in Galaxy-B by model-fitting (Fig.~\ref{fig:galfit_img}) the intensity profiles using the software GALFIT \citep{peng02,peng10}. The H-band (F160W) image is used since it has the highest signal-to-noise among the available bands. We find that fitting and subtracting a double S\'{e}rsic  profile (with indices $\sim 0.5$) gives the best result, assessed by the maximum fluctuations in the residual image (found to be $< 1 \sigma$). Care is taken to simultaneously subtract all nearby sources within $\sim 8^{\prime\prime}$. We also fit the same profile, convolved with the appropriate point-source-functions (PSFs), to the images in all other observed wavelength windows to ascertain consistency. We find no $> 1 \sigma$ fluctuations in the respective residual images. We find that the two S\'{e}rsic  profiles for Galaxy-B are clearly not spatially coincident with the sub-millimeter bright compact core (discussed below in Sec.~\ref{sec:submm_fit}), but rather fall $\sim 0.8^{\prime\prime}$ on either side of it (and hence with a spatial separation of $\sim 0.16^{\prime\prime}$ between the two S\'{e}rsic  profile centres). We also find remarkable consistency of the axes ratio and position angles of the two S\'{e}rsic profiles, hence likely part of a continuous stellar structure. We repeat this procedure for Galaxies A and C. Galaxy-A is found to have a complex morphology which is fit with a double S\'{e}rsic in differing (almost orthogonal) orientations. Galaxy-C on the other hand is found to resemble Galaxy-B in terms of an accordance with two co-aligned S\'{e}rsic  profiles with low indices ($<1$) on  either side of sub-millimeter core. But in this case the separation is $\sim 0.2^{\prime\prime}$, making it more compact than Galaxy-B. 

To measure integrated effective radii ($\rm r_{e}$) of each of the galaxies, we did an additional round of fitting using single S\'{e}rsic profiles. Besides using the generic version of the S\'{e}rsic profile, we also invoke the $\rm m=1$ Fourier component to account for the lopsidedness of the stellar components. This is due to quantify the emission asymmetry we observe in each of the cases which is later discussed in Sec.~\ref{sec:lopsided_outsidein}. We also measure growth curves using  circular apertures with their centers fixed at the ALMA core of the galaxies to independently determine the effective radius. For Galaxy-A, we only use the northern component of its complex near-IR double emission, which we determine as the primary stellar region due to its proximity to the sub-mm emission.
We use the average of all three methods that are consistent within $15-20\%$, to obtain the final measurements reported in Table~\ref{tab:1}: $3\pm 1 \,\rm kpc$, $9\pm2 \,\rm kpc$ and $4 \pm 1\,\rm kpc$ for Galaxies-A, B and C respectively. It is noteworthy that for each galaxy, the single profiles are characterised by S\'{e}rsic indices $<1$, indicating disk-like morphology (and hence we refer to them as disks throughout this work, a conclusion which will be reinforced in Sec.~\ref{sec:morph_character}).  The ellipticities measured for the three galaxies are $0.5 \pm 0.1, 0.15 \pm 0.05$ and $0.5 \pm 0.2$ (A, B and C), indicating an almost edge-on orientation for Galaxy-B.
\begin{table*}
{
\caption{\centering Derived morphological Spergel parameters in sub-millimeter}
\label{tab:2}
\centering
\begin{tabular}{cllll}
\hline\hline
ID                        &                       & A             & B              & C   \\
\hline
effective radius$_{\rm major}^{**}$       & arcsec             & $0.092 \pm 0.004$         & $0.148 \pm 0.002$          & $0.122 \pm 0.054$  \\
       & kpc             & $0.72 \pm 0.02$         & $1.15 \pm 0.01$          & $0.95 \pm 0.42$  \\
Ellipticity       &             & $0.62 \pm 0.03$         & $0.40 \pm 0.01$          & $0.47 \pm 0.22$  \\
Position angle       & degrees            & $-90.6 \pm 0.6$         & $-27.4 \pm 0.3$          & $-46.4 \pm 0.6$  \\
$\Delta$angle$_{disk,bulge}$ (1)      & degrees            & \ \ \ \ \ \ --        & $3.4 \pm 1.5$          & $2.9 \pm 1.5$  \\
Spergel index  (2)     &             & $-0.6 \pm 0.2$         & $0.5 \pm 0.3$          & $-0.5 \pm 0.3$  \\
\hline
\hline
\end{tabular}\\
}\vspace{0.2in}
{Notes: (1) The difference between position angles of the stellar disk in HST F160W and that of the sub-millimeter core. The lack of a value in case of Galaxy-A is due to its stellar emission having a complex morphology without a clear position angle for the whole galaxy. (2) Spergel index values 0.5 and -0.6 correspond to S\'{e}rsic indices 1 and 4 respectively. $^{**}$ $\sim$ effective radius of the corresponding S\'{e}rsic profile, as has been determined through modelling S\'{e}rsic and Spergel profiles (Tan et al. in prep.) }
\end{table*}

\begin{figure}
\centering{
\includegraphics[width=0.5\textwidth]{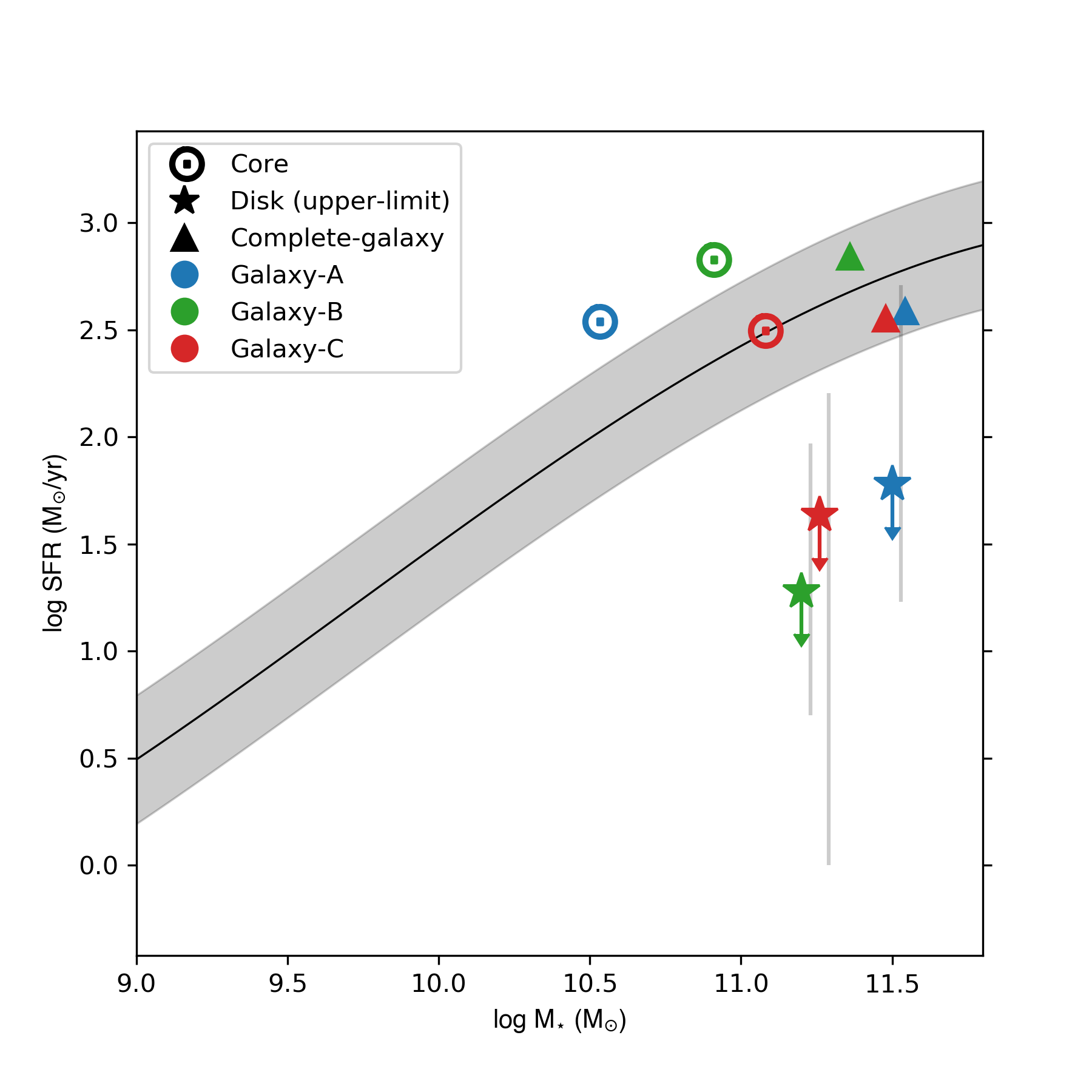}}
\caption{Components of the three galaxies on the star-forming main-sequence: star-bursting bulges and quiescent disks. The SFR vs stellar-mass values of the star-bursting core, the quiescent stellar disk and the combination of the two placed with respect to the star-forming main-sequence \citep{schreiber15} at $\rm z = 3$, shown as the black solid line. The shaded region demarcates the 0.3 dex uncertainty in the relation. We use the $3 \sigma$ ALMA upper-limits as the value for SFR with the associated grey error-bars showing the range allowed by our SED model fitting.The latter have been artificially offset in the x-axis to emphasize the difference in the method of measurement.} \label{fig:galb_MS}
\end{figure}

\subsection{{Sub-millimeter emission analysis}}
\label{sec:submm_fit}

\begin{figure}[!ht]
\centering
\includegraphics[width=0.5\textwidth]{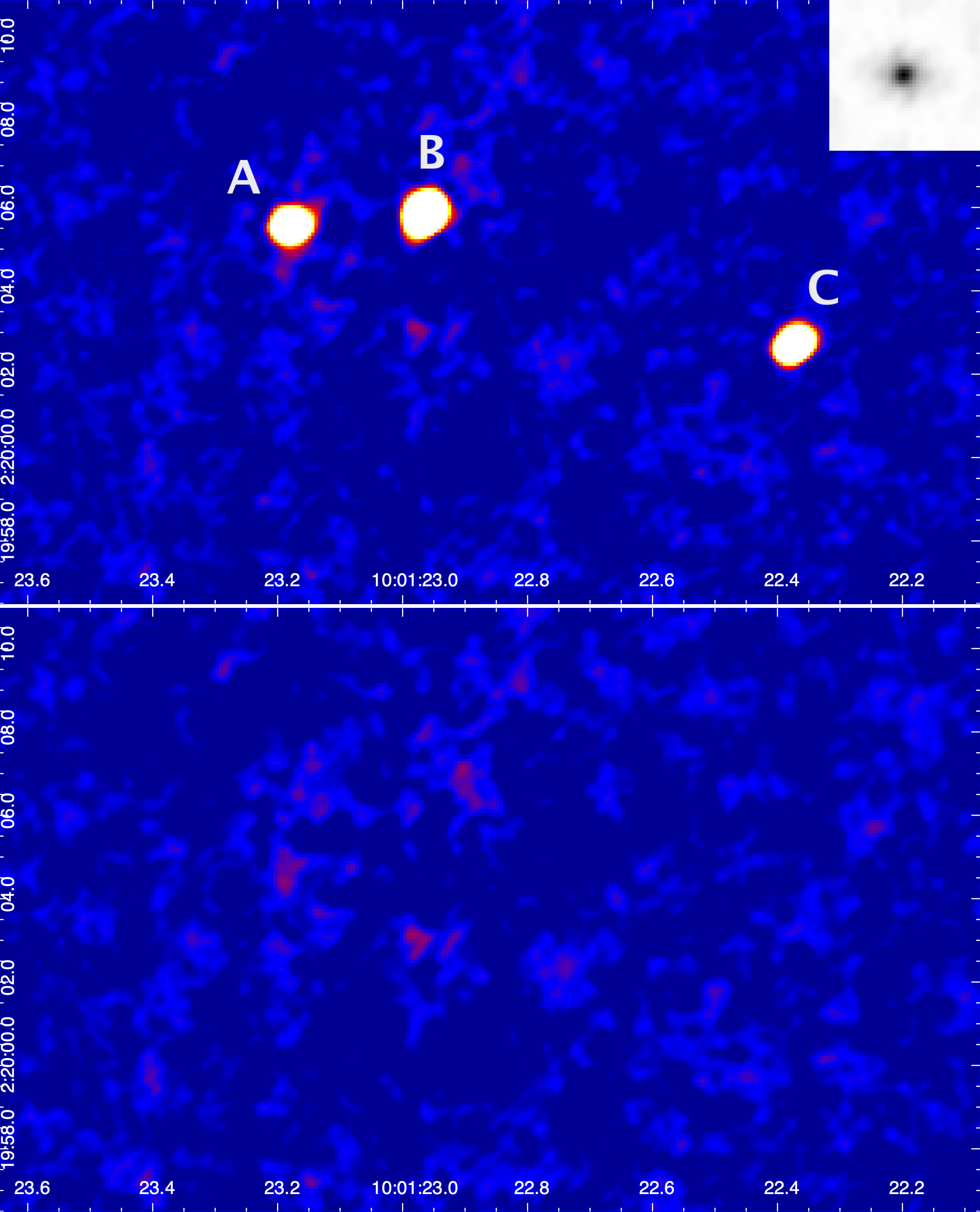}
\caption{The top panel shows the ALMA $870\,\mu\rm m$ image of the core of RO-1001 with the three star-forming galaxies clearly visible. Each of them were fit with intensity profiles which were then subtracted to obtain the residual image displayed in the bottom panel. The ALMA PSF at $870\,\mu\rm m$ is shown in the inset on the top right.} \label{fig_alma_fit}
\end{figure}

The compact highly star-forming regions of the galaxies are detected thanks to their dust emission using ALMA at $870\,\mu$m. It is noteworthy that due to a well characterised PSF and high signal-to-noise, we have the ability to map structures much smaller than the integrated ALMA PSF \citep{rujopakarn19}. We begin with a detailed morphological analysis of the emission region in Galaxy-B. Due to the data being distributed over multiple partially overlapping pointings, we combine the ones with Galaxy-B within their primary beam. To do so, we follow the procedure outlined in \citet{gomez-guijarro21}, for combination and stacking of individual pointings in ALMA. We phase shift each of them to set the source coordinates at their respective phase center. Finally, the pointings are combined into a single measurement set which provides the maximum sensitivity available with our data.	 

Since the  S\'{e}rsic  profile,  commonly used to study galaxy morphologies, cannot be easily extended to the UV space due to its Fourier transform  not being analytically expressible, we use a substitute. We exploit the Spergel profile \citep{spergel10}, recently incorporated within the  MAPPING procedure of GILDAS. This has been found to correlate well with the S\'{e}rsic  profile allowing us to extract S\'{e}rsic  parameters corresponding to the Spergel parameters we determine. To also determine the robustness of the fitting procedure (Fig.~\ref{fig_alma_fit}; Table.~\ref{tab:2}), we perturb the initial parameters within a factor of 5 over a total of $\sim 200$ times and repeat the fitting. We also artificially inject $1000$ times the best fitting model of each galaxy at different empty locations (one at a time) and  fit them individually to estimate reliable error for our measurements based on the a-posteriori dispersion of resulting parameters. This procedure also shows that there are no detectable systematic biases in the measured quantities. 

For Galaxy-B, we obtain an almost-perfect exponential profile characteristic of a disk of effective radius $\sim 1.1\,\rm kpc$, with a Spergel index of $0.55 \pm 0.12$ (S\'{e}rsic index $\sim 1$) and an ellipticity of $0.40 \pm 0.01$. From the model hence returned, we measure the $870\,\mu$m observed-frame flux. This is converted to a total-infrared luminosity ($\rm L_{\rm IR}$; $8-1000\,\mu$m) using the integrated SED from the co-added photometry of RO-1001 \citep{daddi21}. This can then be converted to a SFR of $674 \pm 106\,\rm M_{\odot}\,yr^{-1}$ using the widely adopted conversion relation \citep{kennicutt98, daddi10b}:
\begin{equation}
\rm SFR_{\rm IR} [\rm M_{\rm \odot}\,yr^{-1}] = 1.0 \times 10^{-10} \, \rm L_{\rm IR} [\rm L_{\odot}]
\end{equation}
The error for this measurement is determined from the $1\sigma$ dispersion in the dust-temperature dependent parameter $<\rm U>$, that is usually invoked during the conversion of the $870\,\mu$m flux to $\rm L_{\rm IR}$. We use this to account for variations in dust temperature of individual galaxies with respect to the average value. This approach of evaluating SFR uncertainties is over-conservative, as it does not take into account that the integrated bolometric IR luminosity (hence SFR) of the 3 galaxies that is known to better than $10\%$ accuracy, given the well sampled IR SED including Herschel \citep{daddi21}.

In Galaxy-A, we obtain a much different Spergel index ($-0.61 \pm 0.04$) that translates to a S\'{e}rsic  index $\sim 4$. The situation is similar in Galaxy-C with a Spergel index of $-0.48 \pm 0.12$, although it has an additional point source adjacent to it (at $0.14^{\prime\prime}$). The sizes of these profiles are very compact with effective radii $\sim 0.7\,\rm kpc$ and $0.9\,\rm $kpc. We measure the SFR using the same prescription as that for Galaxy-B, and estimate values of $345 \pm 55\,\rm M_{\odot}\,yr^{-1}$ and $313 \pm 50\,\rm M_{\odot}\,yr^{-1}$ for Galaxies A and C respectively.    

\subsection{Determining the quiescence of the stellar regions}

Besides the measurements for the sub-mm bright compact cores, the high resolution $870\,\mu\rm m$ map (Sec.~\ref{sec:obs_data}) also provides us with constraints on the levels of star formation in the surrounding regions. We are mainly interested in the estimates for the extended stellar regions detected in near-IR, that lack any sub-mm emission. Using the H-band models as fixed priors, we measure $3 \sigma$ upper-limits for each of the galaxies by simultaneously fitting them along with the primary profiles for the cores in the ALMA data (uv plane). For Galaxy-B, we estimate a $3 \sigma$ flux upper-limit of $0.25\,\rm mJy$ (corresponding to an SFR upper-limit of $19\,\rm M_{\odot}\,yr^{-1}$) for the stellar disk, while we obtain $0.77\,\rm mJy$ and $0.55\,\rm mJy$ ($60\,\rm M_{\odot}\,yr^{-1}$ and $43\,\rm M_{\odot}\,yr^{-1}$) for galaxies A and C respectively. This provides clear evidence of almost all star formation being concentrated at the core leaving the massive stellar disks devoid of it and well below the star-forming main-sequence (Fig.~\ref{fig:galb_MS}). 

We note that the estimation of the star formation upper limits in the stellar disks should be considered as overly conservative. This is because the conversion to SFR is done based on the average IR SED in the group which is driven by the co-addition of the emission from all three sub-mm bright galactic cores in RO-1001 that have a much higher  SFR surface density ($\Sigma_{\rm SFR}$; e.g., see Fig.~\ref{fig:galb_model}). Hence, the dust temperature in the disks would likely be lower due to a softer radiation field \citep[based on the star-formation surface density vs. dust temperature relation,][see also \citealt{magdis12,magnelli14, daddi15}]{valentino20b}, which would in turn drive our $3 \sigma$ upper-limit even lower, probably at least by factors of 2--3. 

\subsection{{Star formation history modelling}}
\label{sec:SED_fit}

\begin{figure*}[ht]
\centering{
\includegraphics[width=\textwidth]{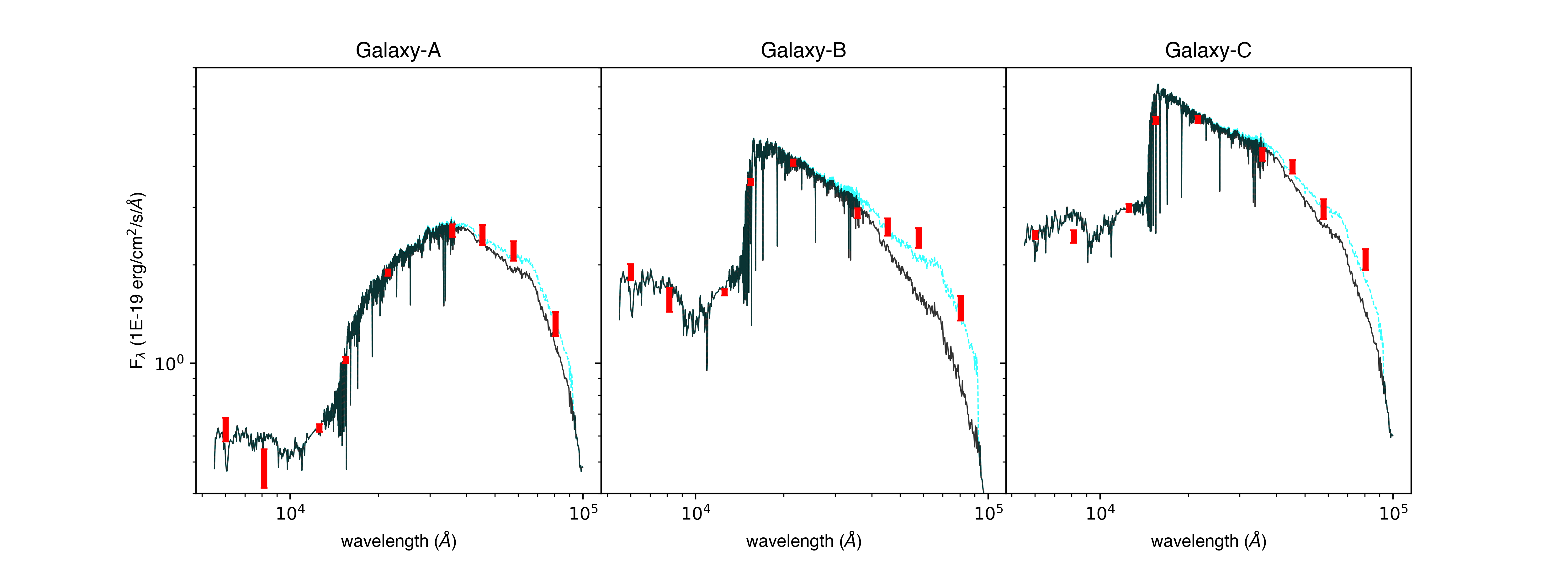}}
\caption{The best-fit SED models in optical and near-IR for each of the three star-forming galaxies in RO-1001. The photometry with their respective errorbars are shown in red. The cyan contours show a modified SED with an addition of an example $0.1\,$Gyr old highly obscured ($\rm A_{V} = 6.5-7.5$) constant star-formation history SED in each case. In each case, the resulting stellar-masses are in agreement with our $\rm CO[3-2]$ estimates within 0.1 dex and are hence a representation of contributions of the star-bursting cores.} \label{fig:sed}
\end{figure*}  

\begin{figure}[ht]
\centering{
\includegraphics[width=0.495\textwidth]{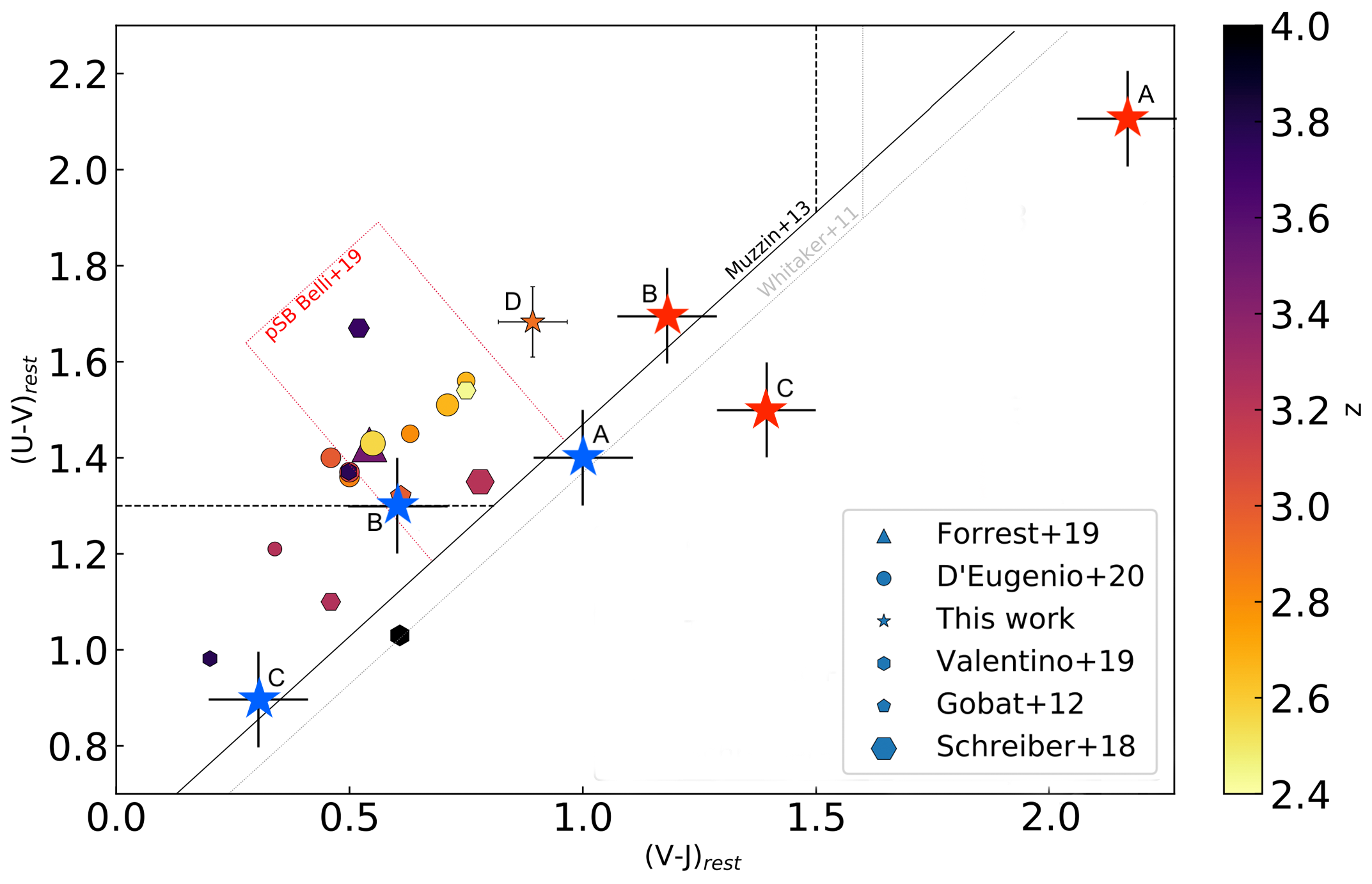}}
\caption{The UVJ colors of high-z QGs ($\rm z>2.5$) adapted from \citet{deugenio20a}. The post-star burst region is demarcated with the red box as specified in \citet{belli19}. The marker sizes are proportional to the stellar mass. The red and blue stars are the dust uncorrected and corrected measurements of the quiescent stellar disks in the three star-forming galaxies within our sample. The final star in orange marks the location of Galaxy-D, a quiescent galaxy photometrically confirmed to be part of RO-1001 \citep{kalita21b}} \label{fig:uvj}
\end{figure}

We make use of the photometry obtained in observed frame optical and near-IR, to estimate a star formation history (SFH) and thereby the age of the stellar population as well as stellar masses. The measurements can primarily be attributed to the stellar disks since the prodigious sub-millimeter flux from the cores is by design at the expense of UV rest-frame flux. However, during our fitting procedure, we do fix an additional core component in each case using positional priors obtained in Sec.~\ref{sec:submm_fit}. Except in the IRAC bands, all others return fluxes $<5\%$ of the total galaxy flux. However for IRAC, the PSF is a factor $\sim 10$ larger than that of HST and a factor $\sim 5$ compared to Ultra-VISTA. Hence in this case, the sources are unresolved and this leads to the photometry corresponding to the integration over the whole galaxy. Albeit, in each of the cases the stellar mass within the core is expected to be $<10\%$ (based on estimates discussed in Sec.~\ref{sec:rotation_gas}) that ensures a minimal contribution. Nevertheless, Fig.~\ref{fig:sed} does suggest progressively higher fluxes in the IRAC bands, leading to offsets from SED models. To limit the influence of the rapidly star-forming core, primarily in the IRAC bands with insufficient resolution to disentangle the core from the disk, we execute the SED fitting in three stages. The first is done using photometry up to the bands where the galaxies are well-resolved ($2.1\mu$m). We then include the remaining 4 bands in pairs of two. We find that the results from the first two stages are almost identical within their errors and hence use the results of the second stage (up to the $4.5\mu$m band; due to better constraints). We show in Fig.~\ref{fig:sed} that the net photometry at the third stage is reproducible using the SED derived at stage-two and an additional highly obscured component for the core.

For a description of the SED model fitting, we refer the readers to Appendix~\ref{sec:sed_info}. The procedure for Galaxy-B returns the lowest reduced $\chi^{2}$ of $2.0$ for the two declining SFH models used (composite-$\tau$ and delayed-$\tau$). In case of a constant star-formation model this value is found to be $6.8$ that translates to a $\Delta \chi^{2} = 38$, making it highly unlikely. Hence, this indicates that the disk in Galaxy-B is undergoing a decline in star-formation, as already conclusively demonstrated by the ALMA upper limts. We also determine this decline to be significant by measuring the age/$\tau$ ratio from the delayed-$\tau$ models and it is found to be $7 \pm 1$. Given that the composite-$\tau$  models are relatively more robust, we use them to measure the look-back time of half-mass formation ($\rm t_{50} = 0.5^{+0.7}_{-0.2}\,\rm Gyr$), that can be considered to be the approximate epoch at which the last major star formation episode had occurred. However, the results from delayed-$\tau$ models are also in agreement. Furthermore, we obtain a SFR estimate of $42^{+51}_{-37}\,\rm M_{\odot}\,yr^{-1}$, consistent with the ALMA $3\sigma$ upper-limit. 

The picture is slightly less strict in Galaxy-C followed by Galaxy-A. In the former, the declining models still give the best fit (reduced-$\chi^{2} = 1.9$) and a $\rm t_{50} = 0.2^{+1.3}_{-0.1}\,\rm Gyr$. But the constant star formation model is relatively less unlikely than for Galaxy-B, with a reduced-$\chi^{2}$ of 3.7. Galaxy-A on the other hand has values of 0.8 and 1.1 for almost equally plausible declining and constant star formation models, allowing for possible ongoing star formation in the disk. Nevertheless, the SFR in both these cases ($94^{+66}_{-94}\,\rm M_{\odot}\,yr^{-1}$ for Galaxy-A and  $66^{+446}_{-49}\,\rm M_{\odot}\,yr^{-1}$ for Galaxy-C) are still in general agreement with the much more stringent ALMA $3\sigma$ upper-limit. 

We also derive  UVJ colors (Fig.~\ref{fig:uvj}) not so much to  ascertain quiescence that is already unambiguously demonstrated by ALMA, but rather to compare them to more classical exemples of high redshift quiescent galaxies. Of special interest are the widely studied post-starburst (PSB) galaxies that have experienced quenching of star-formation over the last $< 0.8\,\rm Gyr$ and generally show low dust attenuation \citep{gobat12, schreiber18, forrest20, deugenio20a, forrest20, valentino20}. The PSB ages are potentially similar to the mass-weighted ages derived for our sample, while attenuation in our case is larger (Table.~\ref{tab:1}). 
We find that the observed UVJ colors of our quiescent disks are much redder than those of known population of quiescent systems at $z\sim3$. Also, at least in two cases they are scattered just outside of the formal UVJ-passive boundary. However, when correcting for 
dust reddening the colors of the three galaxies in RO-1001 more closely resemble those of PSBs as well as of the massive quiescent galaxy with a mass-weighted age $>1\rm\,Gyr$ \citep{kalita21b} within the same group (Fig.~\ref{fig:uvj}).

Finally for Galaxy-B, we pay special attention to the spatial decomposition of the stellar mass estimates (listed in Table.~\ref{tab:1}) derived from the SED fitting procedure -- by obtaining separate photometry of the northern and southern sections of its elongated disk. This is to quantify the lopsidedness in the stellar structure. We hence find the total stellar mass of $1.6 \pm 0.4 \times 10^{11}\,\rm M_{\odot}$ for the complete stellar region is divided into $1.1 \times 10^{11}\,\rm M_{\odot}$ and $0.5 \times 10^{11}\,\rm M_{\odot}$ for the northern and southern segments. It is noteworthy that for this analysis, we only use observations up to the $2.1\,\mu$m band beyond which the resolution is too low to spatially disentangle the two parts. 

\subsection{Rotating gas traced by $\rm CO[3-2]$ transition.} 
\label{sec:rotation_gas}
\begin{figure*}[ht]
\centering{
\includegraphics[width=0.63\textwidth]{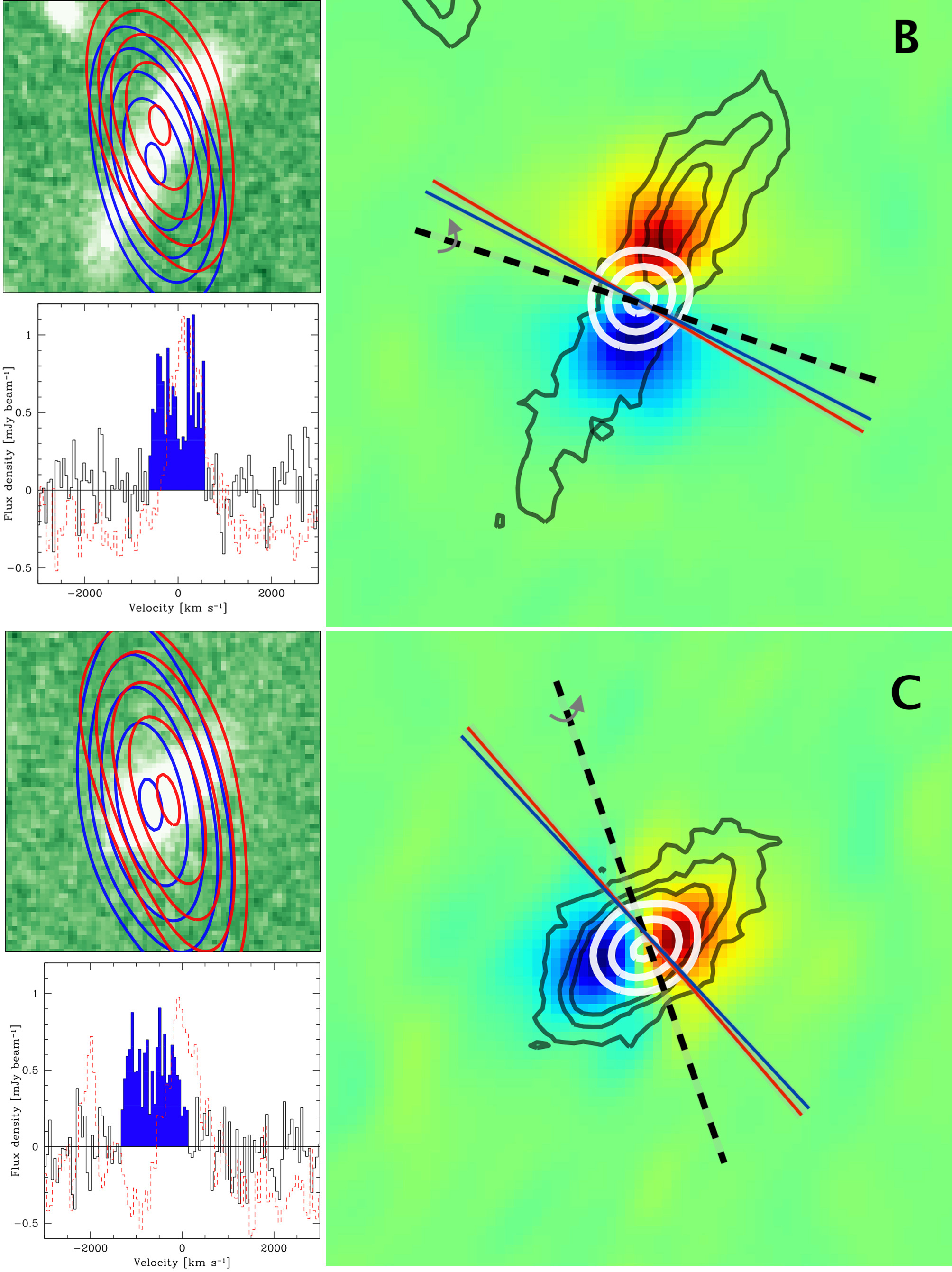}}
\caption{Two panels showing the rotational properties in Galaxies B and C with three individual sub-images: (top-left) the NOEMA $\rm CO[3-2]$ emission profiles for the \textit{blue} and \textit{red}-shifted sections with respect to the average galaxy profile shown as contours superimposed on the F160W image; (bottom-left) The NOEMA spectra with the $\rm CO[3-2]$ emission line in blue and the local Lyman-$\alpha$ profile in red; (right) The contours of the stellar disk (black; beginning at $8\sigma$ with increments of $4\sigma$), the compact core (white; beginning at $100\sigma$ with increments of $50\sigma$ for Galaxy-B and at $50\sigma$ with increments of $25\sigma$ for Galaxy-C) and a representation of the \textit{blue} and \textit{red}-shifted $\rm CO[3-2]$ emission locations with the size of the compact core as seen in ALMA. The expected rotational axes have been also been provided for each component -- disk in red, core in blue and $\rm CO[3-2]$ emission as black dashed lines. The NOEMA spin axes uncertainties are large enough to make the difference with the other component not statistically significant.} \label{fig:rotation_galaxies}
\end{figure*}
A direct spectroscopic tracer of the gas content of the galaxies is available in the form of NOEMA $\rm CO[3-2]$ transition observations. For Galaxy-B, we use its integrated luminosity ($\rm log(L^{'}_{\rm CO(3-2)}/({\rm K\,km^{-1} pc^2})) = 10.46$) to estimate a gas mass. This is calculated from the line flux \citep{daddi21} using the relation prescribed in a recent work \citep{silverman18}. Using a standard $\rm R31=0.5$ (Daddi et al. 2015; Valentino et al. 2020; Boogardt et al. 2020), we hence derive $\rm log(L^{'}_{\rm CO(1-0)}) = 10.76$. The gas mass is finally computed as M$_{\rm mol} = \alpha_{\rm CO} \times {\rm L}^{'}_{\rm CO(1-0)}$ by adopting a conversion factor $\alpha_{\rm CO} = 0.8 \ {\rm M}_{\odot}{\rm (K\,km s^{-1} pc^2)}^{-1}$ as expected in starbursts (given location of the cores, with respect to the star-forming main-sequence in Fig.~\ref{fig:galb_MS}, and their very high SFR surface densities). This provides values of $4.6 \times 10^{10}\,\rm M_{\odot}$, $0.7 \times 10^{10}\,\rm M_{\odot}$, $4.2 \times 10^{10}\,\rm M_{\odot}$ for Galaxy-B, A and C respectively. Although, we do acknowledge that this estimate is dependent on our choice of conversion factors. Nevertheless, we can place a strict upper-limit on the gas mass using the dynamical mass estimates calculated below (i.e., in the limiting case in which there was no stellar mass nor dark matter in the cores), which is less than $2 \times$ above the current values in each case. A Milky-Way-like $\alpha_{\rm CO}$ would have resulted in a factor of 5 higher gas-mass. Hence, these are in direct conflict with the dynamical mass values. Moreover, any increase in gas-mass would also indicate almost a negligible stellar mass, which in turn would place the galaxy-cores at a much higher offset from the star-forming main-sequence. This would suggest even more star-burst-like characteristics.

The width of the emission line in Galaxy-B is used to determine the dynamical mass within the dusty core where the gas is expected to be confined. Given the sub-millimeter emission profile of the core is found to resemble a disk, we measure the $\rm M_{\rm dyn,r_{e}}$ within the effective radius $\rm r_{e}$ using the following relation \citep{daddi10}:
\begin{equation}
\rm M_{\rm dyn,r_{e}} = 1.3 \times \frac{\rm r_{e} \times \left(FWZV_{\rm CO[3-2]}/2\right)^2}{G\,{sin^{2}}i} \pm 12.5\%
\end{equation}
The effective radius returned from the sub-millimeter emission profile is used in this equation. The width of the line, quantified by the parameter $\rm FWZV_{\rm CO[3-2]}$ is already reported in Table.~\ref{tab:1}, while the inclination angle (i) is determined based on the ellipticity of the sub-millimeter profile. Subsequently, we measure the total dynamical mass, $\rm M_{\rm dyn,tot} = 2\times \rm M_{\rm dyn,r_{e}} = 1.3 \pm 0.1 \times 10^{11}\,\rm M_{\odot}$. The factor of 2 in invoked since only half of the mass is contained within $\rm r_{\rm e}$. We use the same procedure to obtain $\rm M_{\rm dyn,tot} = 2.6 \pm 0.3 \times 10^{10}\,\rm M_{\odot}$ and $1.7 \pm 0.2 \times 10^{11}\,\rm M_{\odot}$ for Galaxies A and C. However, given their cores have morphology closer to a S\'{e}rsic index of 4, we re-estimate the $\rm M_{\rm dyn,tot}$ using the following formula applicable to an elliptical galaxy dominated by random motion.   

\begin{equation}
\rm M_{\rm dyn,r_{e}} \approx 5.0\, \frac{\rm r_{e}\,\sigma_{e}^{2}}{G}
\end{equation}

We measure the total dynamical masses of $\sim 2.2 \times 10^{10}\,\rm M_{\odot}$ and $2.4 \times 10^{11}\,\rm M_{\odot}$ for Galaxies A and C respectively, reasonably consistent with our earlier estimates. Although the relatively high discrepancy in case of Galaxy-C likely points to the rotational-dominance expected in a gas-rich structure. Nevertheless, we do not expect the structures to fully conform to such a dynamical arrangement due to the inherent dissipative nature of gas within them.

Finally, two separate continuum images are created for Galaxies B and C, by dividing the whole spectral range into two segments that are separated by the average location of the $\rm CO[3-2]$ emission of the whole galaxy. The offset between the peaks trace the plane of rotation, which are found to be in agreement with the major axes of the respective galaxies in both sub-millimeter and near-IR (Fig.~\ref{fig:rotation_galaxies}). A similar analysis is not possible for Galaxy-A due to a low signal-to-noise of the $\rm CO[3-2]$ detection. 

\section{Results}
\label{sec:disc}

\subsection{Compact star-forming cores in extended quiescent stellar disks}

The spatial decomposition of surface-brightness profiles of the galaxies reveals extremely compact sub-millimeter bright rotating cores and extended near-IR detected co-planar stellar disks showing varying levels of lopsidedness (Fig.\ref{fig:RO-1001}, right panel). The size contrast is evident from the high near-IR to sub-mm radius ratios: $\sim5$--8 for Galaxy-B, while in case of Galaxy-A and C it is $4\pm1$ and $5\pm1$. Majority of the star formation is concentrated in the cores, with the IR-based SFR in the cores being a factor of $6-36 \times$ higher than the $3\,\sigma$ upper-limits in the disks. The mass concentrations also displays this dichotomy between the cores and the stellar components. Using the difference between the $\rm CO[3-2]$ derived dynamical masses and the gas masses, we obtain approximate stellar masses of the cores for each galaxy ($\approx 2 \times 10^{10}\,\rm M_{\odot}, 8 \times 10^{10}\,\rm M_{\odot}$ and $12 \times 10^{10}\,\rm M_{\odot}$). These are concentrated within the areas $> 60\, \times$ compact than the extended stellar regions that feature stellar masses of $3.2 \pm 1.3 \times 10^{11}\,\rm M_{\odot}$, $1.6 \pm 0.4 \times 10^{11}\,\rm M_{\odot}$ and $1.8 \pm 0.5 \times 10^{11}\,\rm M_{\odot}$.

\subsection{Morphological characterisation}
\label{sec:morph_character}
\begin{figure*}
\centering{
\includegraphics[width=\textwidth]{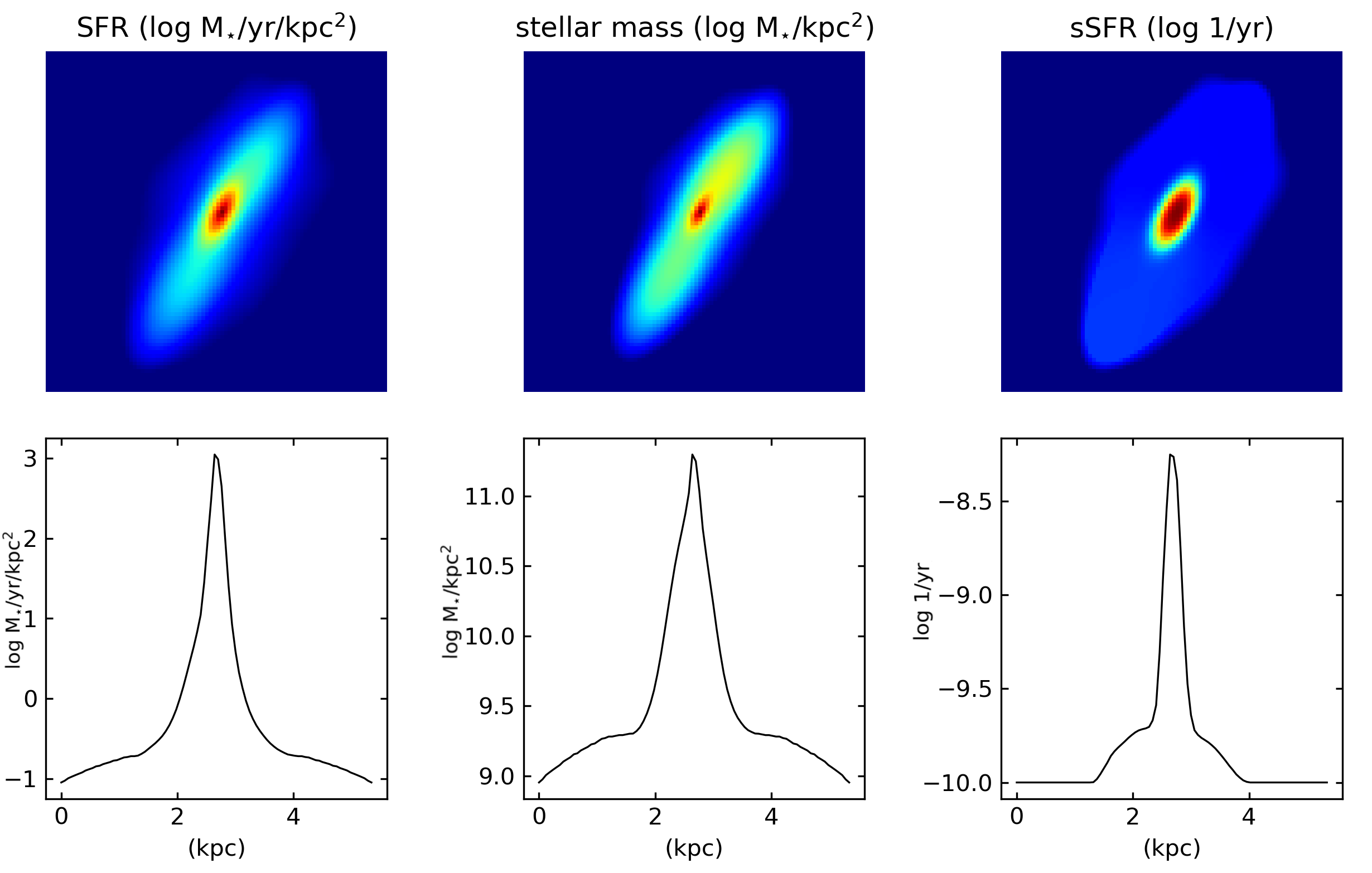}}
\caption{Spatial distribution of Galaxy-B stellar mass and star formation rates. (Left) The upper panel shows SFR distribution model of Galaxy-B, with the morphology determined by the F160W stellar emission in case of the disk, and the ALMA dust emission for the core. The amplitudes are determined by the SFR of the two components. The lower panel is a 1-D representation of the plot above, after a rotation the make the major axis horizontal.  (Middle) The stellar mass distribution following the same procedure. Here the mass in the core is determined from the difference between the dynamical mass and gas-mass determined from the $\rm CO[3-2]$ emission line. (Right) The specific SFR provided by the ratio of the SFR and the stellar mass.} \label{fig:galb_model}
\end{figure*}

The single-peaked profiles point to these galaxies consisting of singular (albeit, lopsided) disks with central cores. This is corroborated by the single S\'{e}rsic profile fits (with indices $\lesssim 1$; Sec.~\ref{sec:morph_stel_fit}) to their near-IR surface-brightness profiles. About $10-40\%$ of the stellar mass is concentrated in the cores occupying $<1\%$ of the surface area of the whole galaxy. This is illustrated in Fig.\ref{fig:galb_model} for Galaxy-B, the most extended of the three galaxies. We find comparable contrasts in the other two galaxies 
(based on their near-IR to sub-mm ratio, Table~\ref{tab:1}). Hence, the cores of these galaxies are likely forming the future stellar bulges. This is characterised by their nuclear location, masses, the extreme compactness similar to $z\sim2$--3 passive galaxies \citep{daddi05}, and their sub-millimeter surface brightness profiles with S\'{e}rsic index $\sim 4$ for Galaxies-A and C, which is relatively unusual \citep[][ indicate that the indices should be closer to that for a disk]{barro16,martig08}. Galaxy-B although has an index $\sim1$, but the higher ellipticity of the core ($0.40\pm 0.01$) in comparison to the disk ($0.15 \pm 0.05$) still supports the idea of bulge formation.

For Galaxies B and C, we do not expect their stellar disks to have been perturbed by major-mergers involving equal-mass galaxies (with ratios 1:1 to $\sim\,$1:3), which have been studied in multiple simulations over the years \citep[e.g.,][]{bendo00, bournaud05b, nevin19}. Firstly, we conclude a lack of any visible disturbance in the disks from our well resolved and high signal-to-noise HST imaging (Fig.~\ref{fig:RO-1001}, right). Had these undergone major-mergers, they would have had to consist of two perfectly aligned disks at a very early stage of collision (pre-coalescence), the probability of which is $0.1\%$ and $0.4\%$ for Galaxies B and C respectively based on the probability of the specific geometrical orientation (having uncertainties equal to the thickness of the disks). Coupled with the concurrence of both these galaxies are detected in the same exact scenario, the probability reduces to a negligible $0.04\%$. Moreover, the temporal aspect also makes such a scenario even more unlikely -- given that both Galaxy-B and C are detected in the same situation within the observation window. Furthermore, had a major-merger been the case, one would have detected a bi-modality in the mass distribution associated with each participating galaxy rather than a centrally peaked profile, especially seen in Galaxy-B (Fig.\ref{fig:galb_model}). 
An alternate scenario would have been the near-IR detected stellar disk and the sub-millimeter core are actually two separate galaxies. However, we also reject such a scenario due to the sub-millimeter bright segment being perfectly at the geometric center of the stellar emission region in not one but two separate cases within the same group. In Galaxy-A, with its disturbed stellar morphology however, the situation is different from its neighbors. As discussed earlier, the stellar emission can be decomposed into two almost perpendicular orientations. We observe that one of the two (the northern component) is better coinciding with the compact ALMA core, similar to that in Galaxies B and C. We therefore identify it as the primary stellar disk of Galaxy-A. The additional component is likely an in-falling equal-mass galaxy.

In at least two of the galaxies (B and C), the major axes of both the core and the disk are aligned within $\sim 3^{\circ}$, as well as with the major axis implied by the rotational patterns of the gas in the respective cores, determined from the $\rm CO[3-2]$ emission (Fig.~\ref{fig:rotation_galaxies}). This is dependent on a reasonable assumption that the gas is confined within the same area as the dust. Hence, we conclude that these two galaxies have co-planar (and possibly co-rotating) stellar-disks and compact cores. The current data does not allow us to measure the major axis direction of the rotational pattern, or the severely lopsided stellar disk (by a factor $>3$, towards South) of Galaxy-A. For the latter, we are likely observing only the `heavier' part of the disk and hence the disk size and shape parameters cannot be satisfactorily determined. For Galaxy C, the ellipticities of the core and the stellar component are found to be in agreement (within $0.1$). But in Galaxy-B, where the stellar component is observed edge-on (with an ellipticity of $0.15 \pm 0.05$), the core has a larger ellipticity ($0.40 \pm 0.01$). However, it should be noted that all the three cores have values within $0.4-0.6$.


\subsection{Young cores within older disks.} For the three cores, we find short gas-consumption timescales of $20\,\rm Myr$, $70\,\rm Myr$ and $130\,\rm Myr$ and time taken to build the stellar masses, assuming constant rates of star-formation, $\sim 60\,\rm Myr$, $120\,\rm Myr$ and $400\,\rm Myr$. However, the short timescales $\sim 100\,\rm Myr$ for three such neighbouring objects makes them statistically unlikely to be simultaneously observed. Hence we propose that the extreme star formation levels is rather the culmination of an increase over a larger period of time. The rate and the beginning of this increase cannot be ascertained. Nevertheless, these are still similar or smaller compared to the time scales characterising the decline in star-formation of the stellar components, based on the goodness-of-fit from the best-fit SED composite $\tau$-models\citep{schreiber18}: the look-back times to the epoch of half-mass formation ($\rm t_{50}$) are $1.7^{+0.3}_{-0.7}\,\rm Gyr$, $0.5^{+0.7}_{-0.2}\,\rm Gyr$ and $0.2_{-0.1}^{+1.3}\,\rm Gyr$ for Galaxies A, B and C. Hence, the star-bursting cores only started getting built after the disks had already formed about half of their stellar mass and were past their last major episode of star-formation.  

\section{Discussion}

In this section, we revisit the key observational results and discuss how their interplay leads to a coherent picture of the evolution of the three star-forming galaxies in RO-1001. Each of them feature extended stellar disks that are lopsided and quiescent, along with highly star-forming compact cores within them.

\subsection{Lopsidedness and quiescence of the disks}
\label{sec:lopsided_outsidein}

We first begin with the severe lopsidedness of the stellar disks (Fig.~\ref{fig:RO-1001}, right panel). For Galaxy-B, the asymmetry in the stellar emission (of a factor of 2.0 in F160W/\textit{HST}) manifests as a mass lopsidedness of 2.2 (Sec.~\ref{sec:SED_fit}) along the major-axis (Fig.~\ref{fig:residuals}). In Galaxy-C, the emission asymmetry, also along the major axis, is
1.6. The current data does not allow us to measure the major axis direction of the stellar disk of Galaxy-A, but only the lopsidedness upper limit ($>3$) and direction (South), that is nearly perpendicular to the East-West ALMA core major axis (Fig.~\ref{fig:residuals}). 
Such asymmetries, not yet recognised at high redshifts, are in fact a common characteristic among spiral galaxies in the local Universe \citep{sancisi08, rix95, zaritsky97, reichard08}, although relatively less pronounced. Since lopsidedness is usually contributed to tidal interactions and in-situ star-formation asymmetry \citep{sancisi08}, we attribute the strong lopsidedness in our sample to the accretion of material onto the galaxies, i.e. asymmetric gas-accretion \citep{bournaud05} (that is reflected by stellar-mass distribution hence created) and/or the tidal interaction (or `minor-mergers') with in-falling satellite sub-halos \citep{zaritsky97,kazantzidis08,kazantzidis09}, that may also include their assimilation into the primary galaxy halo. These `less-severe' interactions (with mass-ratios from 1:4 or lower) are expected to preserve the stellar disks \citep{bournaud05, hopkins09}. Only in Galaxy-A, can we also associate the imminent major-merger to also be playing a role. 

Meanwhile, the quiescence of the stellar disks (albeit possibly not entirely passive), is the first indication of the presence of an apparent `outside-in' quenching mode in high-redshift massive galaxy populations, in contradiction to the classic `inside-out' configuration \citep{lang14, tacchella15, breda18}. We consider various possible modes of quenching. Since we observe the galaxies to be undergoing an outside-in quenching, this could not have primarily occurred due to feedback from active galactic nuclei (AGN; \citealt{alatalo15}), known to quench galaxies inside-out \citep{tacchella18}. We also find no evidence of AGN activity from X-ray observations \citep{daddi21}, although a radio excess in Galaxy-C is indicative of weak past AGN activity in that galaxy. The process of morphological quenching \citep{martig09}, where the formation of a stellar bulge stabilises the disk against further star-formation, is also improbable. The galaxies are far from being bulge-dominated, essential for this mode of quenching to be applicable, based on their stellar mass distributions. Also unlikely is cosmological starvation \citep{feldmann15} since the availability of gas has already been established in RO-1001. Finally, ram-pressure stripping \citep[RPS;][]{gunn72} could be regarded as a possible contributor as it is known to remove gas from external regions of galaxies \citep{bravo-alfaro00, vollmer01, fumagalli09, boselli14, loni21}, therefore resulting in an apparent outside-in quenching \citep[for a review,][]{boselli21}. It could also lead to compression of the gas in the galaxy which could result in the lopsidedness (although the stellar morphology would not be affected). However, as shown in Appendix~\ref{sec:rps}, a conservative lower-limit of the radius up to which RPS can remove gas from the sample in RO-1001 would be $10-15\,\rm kpc$. This is more than an order of magnitude higher than the kpc-scale cores beyond which the galaxies have suppressed star-formation. Hence we do not expect RPS to be playing a major role.

The only possible scenarios capable of self-sufficiently explaining the observed suppression of star-formation are found to be connected to the aforementioned heavy lopsidedness in the disks. Introduction of such asymmetry, in a differentially rotating disk-like structure induces loss of angular momentum, causing mass to rapidly fall into the central regions \citep{combes85}. This process can be characterised as the triggering of the $\rm m=2$ Fourier component, denoting a bar, by the $\rm m=1$ component that refers to the asymmetry or lopsidedness \citep{bournaud05}, observed in our sample. Bars have already been found to efficiently suppress star-formation up to factors of $\sim 10$ in the disk \citep{khoperskov18}, while driving gas into the core \citep{carles16}, albeit on less spectacular scales than what observed here. This could be a possible case of an early-phase bar-formation, which are usually scarce at high redshifts \citep[e.g.,][]{sheth08}. However, the resolution and sensitivity is not sufficient to confirm it. The accreted clumpy material driving the lopsidedness is also expected to feed the star-forming core, and in the process leading to a stabilization of the disk \citep{dekel09}. Therefore the same outside-in quenching can be expected in all scenarios involving lopsidedness and compact star formation.

Moreover, the expected damping time of the lopsidedness is comparable to the Hubble time \citep{jog09}, explaining their detection throughout our sample and allowing them to act as catalysts of bulge formation under the high redshift accretion conditions. This timescale is not however associated with the gas being funneled into the core, but rather the survival of the asymmetry that drives this process. We propose that accreted material is resulting in the compact star-forming cores within outside-in quenching disks, either directly (mergers) and/or indirectly (lopsidedness funneling gas to the core). The accretion may vary from smooth dust (hence presumably gas) components, to sub-halo stellar `clumps', as in Galaxies B and C (while preserving the disks in the process), or additional massive stellar structures, observed in Galaxy-A.



\subsection{Determining the direction of accretion}
\label{sec:lyman_link}
Due to the confluence of evidence for influence of accretion on the galaxies, we try to discuss possible tracers of this process. In an accreting galaxy, the lopsided part of the disks would be expected to mark the location of impact for the accreted material. Corroborative evidence can be found in our sample when the models of the primary disks in the F160W image and the sub-millimeter bright cores in the ALMA image are subtracted. Residual emission (stars in near-IR emission and dust in sub-mm) is clearly detected in each case (Fig.~\ref{fig:residuals}). In Galaxy-A, its F160W and the associated ALMA residuals demonstrate the presence of the similarly-massive companion merging into the disk. For Galaxies-B and C, their respective F160W residuals are much fainter (by $\sim 3$ orders of magnitude) and are likely in-flowing small-mass `clumps' within or adjacent to the disks.

These are found to be preferentially located in the direction of the lopsidedness of the stellar-disks. We hence suggest that these indicators could be tracing the mass inflow into the core. 

\section{A generalized picture accretion-driven galaxy evolution} 

In this section, we put forward a way to fit our work within the general framework of high-z massive galaxy evolution. Throughout, we suggest how our results indicating accretion onto galaxies is well suited to provide a general picture of galaxy mass-buildup at these redshifts. 

\subsection{Comparison to high-z compact star-forming galaxies}

We start of by making a comparison with the widely studied cSFG population at similar redshifts  \citep{cimatti08, ricciardelli10, fu13, ivison13, toft14, toft17, elbaz18, gomez-guijarro18, gomez-guijarro19, puglisi19, puglisi21}. These cSFGs also feature compact kpc-scale sub-mm bright regions, as is observed in our sample. In fact, \citet{puglisi19} already suggested these regions/cores to be above the star-forming main sequence (as seen in our sample, Fig.~\ref{fig:galb_MS}). However, the surrounding disks being well below the star-forming main sequence as in our sample (Fig.~\ref{fig:galb_MS}), making galaxies appear to be quenching `outside-in', have not yet been recognized. A similar scenario might hold for post-starbust galaxies \citep[e.g.,][]{baron22}. Another apparent difference is the average ratio of the near-infrared and sub-millimeter radii for cSFGs, that is $\sim 3$\ \citep{puglisi21, smercina18, lang19}. Although this is likely a lower limit due to many cSFGs remaining unresolved in large sample studies. It remains thus unclear if examples like the three galaxies in RO-1001 with ratios $4-9$ could also present among cSFGs. Moreover, simply the size of Galaxy-B being a factor of $2.0-2.5$ above the mass-size relation \citep{vanderwel14} for star-forming galaxies at $\rm z=3$, adds to the rarity of our sample, especially with regards to cSFGs. Another characteristic of cSFGs seemingly in conflict with our results is the expectation of major-merger dominance within such samples \citep{elbaz18}. Both in galaxies B and C, the stellar disks appear to be well-preserved (Sec.~\ref{sec:morph_character}). 

We can however provide possible reasons behind these apparent discrepancies. Our high-resolution imaging capabilities in both near-IR and sub-mm for RO-1001 is usually not reached in larger statistical studies of cSFGs. Hence we might have simply been able to resolve the galaxies better to determine their properties. This allowed us to study the lopsidedness, the lack of dynamical perturbations (in galaxies B and C) and the apparent `outside-in' quenching, based on which we draw the conclusions of our work. Although, we can also hypothesize a physical reasoning. Large statistical samples usually comprise of galaxies in the field which do not experience the high levels of accretion of gas and satellites expected within dense gas-rich environments like RO-1001 \citep[accretion rate increases with halo mass;][]{dekel09a}. Hence the cSFGs in the regions with less accretion would be more likely to be results of major-mergers driving gas into compact cores while destroying extended stellar disks. This is further exemplified by our direct visual investigation finding a complete lack of similarly massive galaxies with extended and continuous stellar morphology (like in Galaxy B) in the redshift window of $2.6 < z < 3.2$ in both COSMOS and GOODS-South fields. If this is true, cSFGs with compact star-forming cores and very extended quiescent stellar disks (similar to our sample) might be more easily found in dense accretion-rich regions.

\begin{figure*}
\centering{
\includegraphics[width=0.9\textwidth]{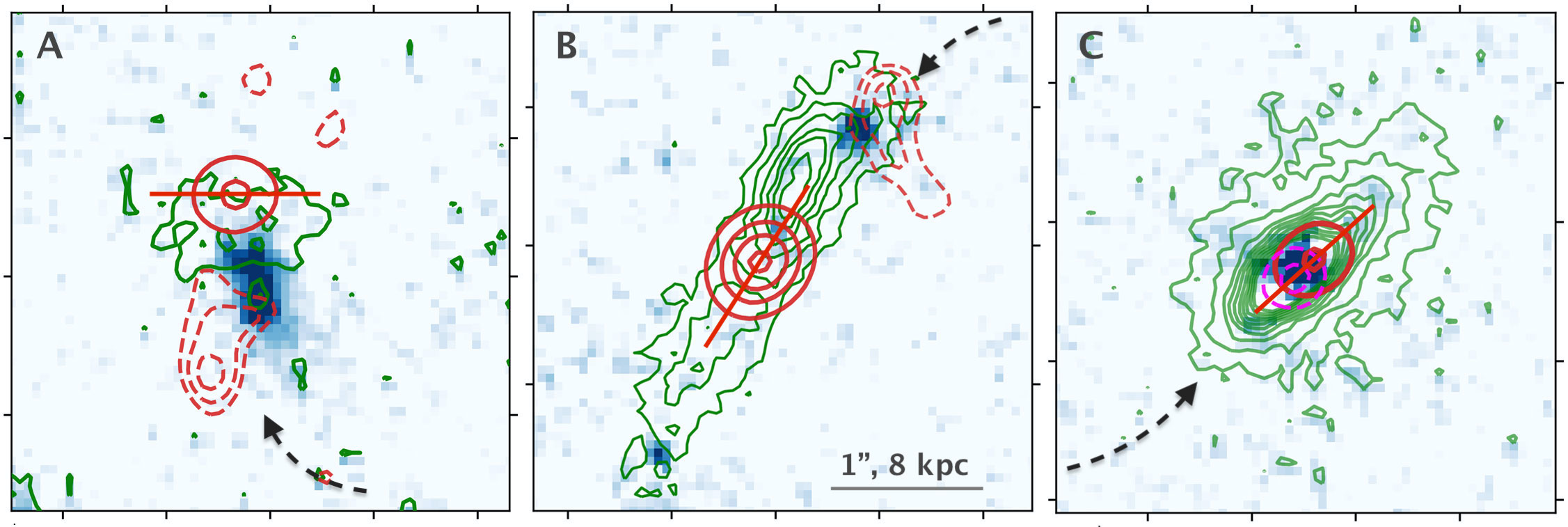}}
\caption{Additional stellar and dust components  tracing the direction of accretion. The three panels showing the three residual F160W images of the galaxies after the subtraction of the primary disks (in green contours; starting at $4\sigma$ with increments of $4\sigma$). The red solid contours show the ALMA $870\,\mu$m emission with the lines starting at $50\sigma$ with increments of $50\sigma$. The red solid lines indicate the respective ALMA major axes. In case of Galaxies A and B, the dashed red contours starting at $3\sigma$ with steps of $0.5\sigma$ displays the ALMA residual emission after the subtraction of the primary emission regions. In case of Galaxy-C, the same contours show the additional point source adjacent to the core that was found during the fitting procedure. In this case, the contours are at $15,25\sigma$. Finally, the arrows indicate the likely directions of gas-accretion onto the galaxies based on the asymmetric distribution of all the aforementioned components, which also coincide with the direction in which the disks are lopsided.} \label{fig:residuals}
\end{figure*}


\subsection{A link to the accretion filaments in RO-1001}

Extending on our hypothesis of accretion onto the three star-forming galaxies in RO-1001, it is imperative to associate the discussed properties with the large scale narrative of accretion for the galaxy-group \citep{daddi21}. Any ongoing accretion in RO-1001 would likely be coupled with the established Lyman-$\alpha$-emitting filaments in RO-1001 \citep{daddi21}. Hydrodynamical simulations predict that massive galaxies evolving under the influence of nearby accretion-streams have their major axes aligned along the stream direction \citep{dubois14,codis18}. We observe a consistency with this prediction in our observations (Fig.~\ref{fig:RO-1001}, right panel). We quantify this by finding the probability for the chance alignment of the major axis of each galaxy with one of the three filaments. The range of allowed angles is approximated using the centre of the galaxies and the edges of the `$-18.0$' contour (corresponding to a surface brightness of $10^{-18}\,\rm s^{-1}\,\rm erg\, cm^{-2}\, arcsec^{-2}$) in Fig.~\ref{fig:RO-1001}, since it is the highest brightness level at which the Ly-$\alpha$ traced filaments are detected. We also ensure that the tentative direction of the filaments, qualitatively reported in \citet{daddi21}, are within this subtended angle ($< \pm 10^{\circ}$). We find total chance probabilities of $13\%$, $17\%$ and $12\%$. Additionally, the probability of all three galaxies being aligned with one of the filaments (which is what we observe) is hence $<1\%$. This alignment, especially in case of disk-like galaxies, is attributable to the process of tidal torquing \citep{codis15} by the massive accretion filaments. However, at the core of this lies the tidal interactions the galaxies experience due to infalling satellites (or minor-mergers). Hence, the apparent alignments should be regarded as an indication of the galaxies accreting clumpy material, which has already been suggested in Sec.~\ref{sec:lopsided_outsidein}.

In each case however, we would still expect a level of `bending' of the accretion streams before the final impact on to the galaxies, revealed by offsets between the direction of accretion (Fig.\ref{fig:residuals}) and the preferred filament (Fig.~\ref{fig:RO-1001}, right panel). We propose a scenario where the Galaxy-A, B and C are being fed by the south-western, northern and south-eastern filaments. The swap of the preferred filament for Galaxies A and C from those in their proximity has been made due to better alignment with the major-axis of the galaxies traced by near-IR and sub-millimeter emissions as well as the direction of accretion traced by the lopsidedness and residual emission (Fig.~\ref{fig:residuals}). Otherwise, for Galaxy-C the accretion direction would have been opposite to direction of approach of the filament. Moreover, there is an additional inconsistency of the line-of-sight velocities of the Lyman-$\alpha$ in the south-eastern filament at the galaxy position \citep{daddi21}. For Galaxy-A, it has a large spatial displacement from the direction of the nearest south-western filament. In each case hence, the actual streams of in-falling material would have needed to drastically bend before the final impact onto the galaxies. Nevertheless, even in our preferred scenario there is still a level of bending of the streams necessary as they approach the inner halos of the galaxies. For Galaxy-A, the major axis (measured here from ALMA) versus the lopsidedness and in-falling material direction are nearly orthogonal. Hence the bending is likely occurring in the line-of-sight. The same could also be true in Galaxy-B, where the northern filament with its reduced length likely has a major component along the line-of-sight. Hence, the in-falling stream would need to bend if it has to agree with the expected direction of accretion along the major-axis, on the plane of the sky (Fig.~\ref{fig:residuals}).

\subsection{A unified picture of accretion-driven evolution}

Based on the results of the three star-forming galaxies in RO-1001, we find evidence of massive disk-like galaxies expected to have secularly evolved featuring compact star-forming regions at their cores. We propose that the massive disks build up lopsidedness under the combined effects of gas and satellite accretion which in turn results in loss of angular momentum of the material. This drives the prodigous star-formation in the central regions. Such a scenario is likely being observed in galaxies B and C. Although, gas can also be driven to the core by much more violent major-mergers \citep{toft17, elbaz18}, but this leads to a destruction of the stellar disks which is only anticipated in Galaxy-C. However, we conclude that the \textit{wet compaction} scenario proposed by simulations  \citep{dekel13, zolotov15, tacchella16}, where compact stellar bulges get rapidly built up as a result of accretion-driven violent disk instabilities, is too extreme to reproduce the observations in RO-1001. Such a compaction simply does not allow for creation of extended and massive stellar disks ($\gtrsim 10^{9.5}\,\rm M_{\odot}$) while in our sample we are observing them with masses $>10^{11}\,\rm M_{\odot}$.   


Finally, the presence of massive stellar disks points to the need for a mode of destruction, since they are not usually observed in dense environments at lower redshifts \citep{bundy10}. Dynamical disturbances caused by mergers or galaxy harassment can easily be invoked for this \citep{toth92,kazantzidis06,bullock09,purcell09,bournaud11}. We might already be observing such a scenario in action in Galaxy-A, with its signs of a forthcoming major-merger. Finally, a possible end product may be the compact quiescent spheroidal galaxy\citep{kalita21b} that has been photometrically confirmed to also be within the group, although it never grew beyond $10^{11}\,\rm M_{\odot}$. It is curious however, that we have stumbled upon a galaxy-group with three massive galaxies close together with recent gas-rich star-bursting cores created within the last few 100 Myr. From each of their observed properties, they seem to have undergone a recent shift in evolution with a preference for the central region. A possible reason has recently been discussed in \citet{noguchi21}, where bulges get rapidly built-up in $>10^{11}\,\rm M_{\odot}$ galaxies within dark-matter halo cores that are accessible to cold-gas filaments \citep{dekel09a}. This limit is related to the dependence of migration timescales of clumpy accreted material (already suggested in our sample by their lopsidedness, as discussed in Sec.~\ref{sec:lopsided_outsidein}) that feed the central star-forming region on the stellar mass of the galaxy. Hence, the galaxies in RO-1001, all of which are $>10^{11}\,\rm M_{\odot}$, could simply be experiencing ideal conditions for a rapid buildup of bulges. It should be noted however that we refrain from estimating the timescale of migration of accreted clumps since such a calculation depends on the relative angular momentum of the inflowing material with respect to the disk. This can result in a range between a few tens to a few hundred Myrs, which is highly uncertain and heavily dependent on assumptions that we are not in a position make with the current data. 

\section{Caveats and future prospects}

\subsection{Further investigations}

In our study, we are still lacking information about certain facets of the three galaxies in RO-1001. A spatially resolved spectroscopic study of the quiescent stellar regions would be extremely useful to constrain their star-formation histories. A lot of work in this direction has been done for massive QGs \citep[e.g.,][]{schreiber18, forrest20, valentino20}, with signatures of rapid quenching of star-formation observed in their samples. Although we do determine rapid quenching of the stellar disks in our sample (Sec.~\ref{sec:SED_fit}), conclusive proof will only be available after obtaining spectroscopic data. 

Furthermore, this will also allow the tracing of the metallicity of the stellar disks. It is hypthesized that accreted gas being metal-poor leads to a fall in the metallicity in the accreting galaxy \citep{lehner13, sanchez13, sanchez14, wotta19}. Hence, RO-1001 could be a test-bed for such theories. Finally, spectroscopy will also provide the kinematics of the stellar disk and determine the level of rotational dominance within the structure. This would be necessary to model the evolution of the galaxies as well as similar galaxies in the future featuring signs of accretion. This is especially of interest since we have suggested the possible prevalence of clumps migrating to the core of the galaxies, driving the high levels of star-formation. The timescales of in-flowing material is coupled with the rotation timescales of the respective disks \citep{dekel09}.   

\subsection{Extending the search to other structures}

Asymmetries in the stellar distribution as well as relative position of the sub-mm bright cores of cSFGs are rather common. However, they are mostly associated with galaxy-galaxy interactions \citep{ elbaz18, cibinel19, puglisi19}. However, we demonstrate that if studied in higher resolution, they might reveal effects of accretion of gas and clumps especially in dark matter halos where cold-gas filaments are expected to survive in spite of the surrounding hot medium \citep[][]{dekel09a}. Hence, deep observations of star-forming galaxies within dense-environments would be instrumental in establishing a generalised picture of massive galaxy evolution in dense environments. A few examples are in fact already available. CLJ1449+0856 \citep{gobat11, gobat13, strazzullo16, coogan18} has three highly star-forming galaxies that showcase clear offsets between their near-IR and sub-mm/radio contours, with one of them also featuring AGN activity \citep[S7-N7 system; ][]{kalita21a}. Using unresolved sub-mm data, they were regarded as mergers. However, reconsideration after obtaining high resolution data might reveal the asymmetries to be due to comparable features as those seen in RO-1001. Another such case is the highly star-forming galaxy, GN20, which is part of a proto-cluster at $\rm z=4.05$ \citep{daddi09, hodge15}. This too features a similar offset and is a confirmed member of a high-z dense environment. Additional sources of interest include galaxies within seven other high-z structures with Ly-$\alpha$ halos suggesting abundance of gas \citep{daddi22b, daddi22a}. Future well-resolved studies of these and additional structures hold much promise.


\section{Summary and conclusions}
\label{sec:summary}
Within a $z=2.91$ galaxy-group of dark matter halo mass of $\sim4\times10^{13}\,\rm M_{\odot}$, we observe three massive ($>10^{11}\,\rm M_{\odot}$ galaxies (Fig.~\ref{fig:RO-1001}). They have a combined star-formation rate of $\sim 1250\,\rm M_{\odot}\,yr^{-1}$, and are studied here with high resolution near-IR (HST) and sub-mm (ALMA) data. Each of the galaxies have extremely compact ({\rm effective radius} $\sim 1\,\rm kpc$) sub-millimeter bright cores where almost all of the star-formation and $10-40\%$ of the stellar mass is confined. They will likely form the future bulges of their respective galaxies. These are embedded in co-planar stellar disk-like structures which are $4-8\, \times$ more extended in radius. The key new results that emerge from our work are the following: 

   \begin{enumerate}
      \item All three galaxies show marked stellar disk lopsidedness of varying degrees (from $1.6$ to $>3.0$). The stellar morphology of 2/3 galaxies (B and C) do not show signs of extreme dynamical disturbances however, hence limiting the role of major-mergers. Although, the third (galaxy A) is likely at an initial phase of collision with an equally massive counterpart. These are first, unique cases of high redshift galaxies in which lopsidedness has been recognised to play an important role. 
      \item Based on ALMA 3$\sigma$ upper-limits, we also determine that star-formation is suppressed in the disks. They are found to be located $>1$~dex below the   star-forming main sequence (whose scatter is 0.3~dex), while the star-forming compact cores are mainly located above it (Fig.~\ref{fig:galb_MS}). These galaxies are among the first known cases of `outside-in' quenching observed in the distant Universe. 
      \item Furthermore, the cores could not have been forming stars beyond earlier than a few $100\,\rm Myr$, which is the same epoch by which the disks had already formed about $50\%$ of their stellar mass. This bi-modality in star formation and stellar age coincides with the apparent outside-in quenching of the galaxies.
      \item We conclude that the lopsidedness is intrinsically connected to gas being driven into the centre of the potential well of the galaxies. The reason of this lopsidedness in at least Galaxies B and C is most likely the combined effect of accreted gas and clumps, the latter of which can be regarded as minor-mergers. In case of Galaxy-A, the imminent major-merger would also play a role. 
      \item The importance of minor-mergers due to accretion is exemplified by the alignment of the major axes of all three galaxies with one of the accretion streams established through Ly-$\alpha$ halo morphology in \citet{daddi21}. This is expected to also result in the lopsidedness as well as any residual stellar and dust emission to converge onto the location of impact of the accretion-stream with the galaxies. We detect these signatures in each of the cases. Hence we make the first direct connection between accretion streams and the morphological transformation of galaxies driving compact star-formation in massive galaxies at high redshifts. 
      \item We show that in the presence of accretion, galaxies can build-up compact highly star-forming cores without the need of major-mergers and therefore preserving their stellar disks. A key feature however is an asymmetry in the stellar distribution, which needs to be studied in high resolution to determine if it is driven by accretion of gas and clumps or by violent interactions like major-mergers. Future studies of accretion-rich high-z dense environments would prove to be crucial.  
   \end{enumerate}

Based on our results, we suggest an evolutionary pathway in which a secularly evolving disk within a dense environment with accretion (gas and clumps), grows substantial mass asymmetry and experiences the creation of a compact star-bursting core at the expense of the star-formation within the disk. This is inevitably followed by some form of disruptive event that leads to the destruction of the quiescent disk \citep{toft17}, leaving behind, and also possibly adding to \citep{elbaz18, puglisi21}, a stellar bulge. However, the ubiquity of such a scenario can only be revealed with more well-resolved observations of massive galaxies at the heart of well studied high-redshift dense environments -- that will likely become more accessible in the upcoming James Webb Telescope era.

\begin{acknowledgements}
First and foremost, we would like to express our gratitude to the anonymous referee for the valuable suggestions. We would also like to thank Gabriel Brammer for assistance with the use of grizli for the HST data reduction. R.M.R. acknowledges support from GO15910. V.S. acknowledges the support from the ERC-StG ClustersXCosmo grant agreement 716762. F.V. acknowledges support from the Carlsberg Foundation Research Grant CF18-0388 “Galaxies: Rise and Death” and from the Cosmic Dawn Center of Excellence funded by the Danish National Research Foundation under then Grant No. 140. This paper makes use of the following ALMA data: 2019.1.00399. ALMA is a partnership of ESO (representing its member states), NSF (USA) and NINS (Japan), together with NRC (Canada), MOST and ASIAA (Taiwan), and KASI (Republic of Korea), in cooperation with the Republic of Chile. The Joint ALMA Observatory is operated by ESO, AUI/NRAO and NAOJ. 
\end{acknowledgements}

%
%

\begin{appendix}
  
\section{Spectral Energy Distribution Model Fitting}
\label{sec:sed_info}
In order to fit SFH models to the galaxies, we use the results of the surface brightness distributions in F160W (which also provides the flux in the band) and then obtain the rest of the photometry using the best-fit models as priors in all other bands. We fit this photometry with BC03 stellar population models \citep{bruzual03} to derive the properties of the three massive galaxies (Fig.~\ref{fig:rotation_galaxies}). The redshift is fixed to the spectroscopic values previously published for the sources \citep{daddi21}. We use FAST++\footnote{https://github.com/cschreib/fastpp} to determine the best fitting SED templates through a $\chi^{2}$ minimisation procedure. We fix the metallicity to the solar value as is expected in massive galaxies and implement the Calzetti dust attenuation law\citep{calzetti00}, with a range of extinction values, A$_{\rm V}=0-5$ in steps of 0.02. Finally, the following SFH models are used:   
\begin{enumerate}
    \item Constant star formation models: We use this model featuring no decline in star formation (it remains constant throughout) to check if a high attenuation star-forming scenario is formally consistent with the available photometry. 
    \item Delayed $\tau$-models: This is an exponentially declining SFH here is characterised as $\propto (t/\tau^{2})\,e^{-t/\tau}$ with a peak of star formation at $t=\tau$. The $\tau$ varies within [100\,Myr, $\rm t_{\rm obs}$] with steps of 0.1\,dex, where $\rm t_{\rm obs}$ is the age of the universe at the redshift of the galaxy.
     \item Composite $\tau$-models\citep{schreiber18}: This is similar to the previous model, however with different timescales ($\tau_{\rm rise}$ and $\tau_{\rm decl}$) for the rising and declining phases separated by the epoch $\rm t_{\rm burst}$. 
\begin{align}
\label{eq:composite_sfh}
\mathrm{SFR}_{\rm base} (t) \propto \left\{\begin{array}{ll}
e^{(t_{\rm burst} - t)/\tau_{\rm rise}} & \text{for $t > t_{\rm burst}$} \\
e^{(t - t_{\rm burst})/\tau_{\rm decl}} & \text{for $t \le t_{\rm burst}$} \\
\end{array}\right.
\end{align}
Both $\tau_{\rm rise}$ and $\tau_{\rm decl}$ are varied within the range of $[10\,\rm Myr,3\,\rm Gyr]$, while for $\rm t_{\rm burst}$ a grid of $[10\,\rm Myr, \rm t_{\rm obs}]$ is used. 
    
\end{enumerate}
   
\section{Prevalence of Ram Pressure Stripping}
\label{sec:rps}

We investigate the effect RPS may have on the three star-forming galaxies in RO-1001. We use the following relation from \citet{domainko06} for the minimum radius up to which gas may be stripped from a galaxy ($\rm r_{\rm strip}$):
\begin{equation}
    \rm r_{strip} = 0.5 \rm\, r_{o} \times ln\left( \frac{G\,M_{star}\,M_{gas}}{\varv ^{2}_{gal}\,\rho_{ICM}\,2\pi\, r_{o}^{2}} \right) 
\end{equation}
Here, the $\rm r_{o}$ is the scale radius of the galaxies, while $\rm M_{star}, M_{gas}$ corresponds to the gas and stellar mass of the galaxies. $\rm \varv_{gal}$ is the relative velocity between the galaxy and the ICM, for which we use the approximate virial velocity of the galaxies in RO-1001 \citep[$500\,\rm km\,s^{-1}$;][]{daddi21}. Finally, the density of the ICM ($\rm \rho_{ICM}$) is determined using \citep{boselli21}:

\begin{equation}
    \rm \rho_{ICM} = 1.15\,n_{e}\,m_{p}
\end{equation}
We use the value of $10^{-3}\,\rm cm^{-3}$ for the electron density ($\rm n_{e}$) as is expected in the core of clusters and groups \citep{vikhlinin09, sun09, pratt10}. 

We hence get an $\rm r_{\rm strip} = 10-15\,\rm kpc$ for the galaxies in RO-1001. However, this should be regarded as a conservative lower-limit since RPS efficiency is expected the steeply decrease at high-redshifts. This is primarily due to the contribution of non-gravitational heating which effectively reduces the ICM content \citep{fujita01}. Moreover, the decreasing efficiency of RPS at high redshifts ($\rm z \gtrsim 2$) is also suggested through simulations in \citep{pfeffer22}. Hence, in each of our sample galaxies, the $\rm r_{\rm strip}$ is found to be a factor of $\gtrsim 2$ larger than the stellar disks.

\end{appendix}

\bibliographystyle{aa}
\bibliography{main.bib}

\end{document}